\documentclass[usenatbib,usegraphicx,usehyperref]{mn2e}
\usepackage[dvips]{graphicx}
\pdfoutput=1
\usepackage{blindtext}
\usepackage{lipsum}
\usepackage{amssymb,amsmath,amsfonts,graphicx,aas_macros}
\usepackage{natbib}
\usepackage{color}
\usepackage{times}
\usepackage[bookmarks,bookmarksnumbered,colorlinks=true, citecolor=blue, linkcolor=black]{hyperref}
\usepackage{float}
\definecolor{cincinnati-red}{RGB}{190,0,0}

\newcommand{\lcdm}{{$\Lambda$CDM}}

\defcitealias{mcmillan}{McM11}
\begin{document}
\setlength{\abovedisplayskip}{10pt}
\setlength{\belowdisplayskip}{10pt}
\title[SIDM halos with baryons]{The impact of baryonic discs on the shapes and profiles of self-interacting dark matter halos}
\author[O. Sameie et al.]{Omid Sameie$^{1}$\thanks{E-mail: \href{mailto:osame001@ucr.edu}{osame001@ucr.edu}}\thanks{NASA MIRO FIELDS Fellow},  Peter Creasey$^1$, Hai-Bo Yu$^1$\thanks{Hellman Fellow},
Laura V. Sales$^1$\thanks{Hellman Fellow},  Mark Vogelsberger$^2$\thanks{Alfred P. Sloan Fellow}  
\newauthor 
and Jes\'us Zavala$^3$  \\
$^1$ Department of Physics and Astronomy, University of California, Riverside, California 92507, USA\\
$^2$ Department of Physics, Kavli Institute for Astrophysics and Space Research, Massachusetts Institute of Technology, Cambridge, MA 02139, USA\\
$^3$ Center for Astrophysics and Cosmology, Science Institute, University of Iceland, Dunhagi 5, 107 Reykjavik, Iceland }
\maketitle
\begin{abstract}

We employ isolated N-body simulations to study the response of self-interacting dark matter (SIDM) halos in the presence of the baryonic potentials. Dark matter self-interactions lead to kinematic thermalization in the inner halo, resulting in a tight correlation between the dark matter and baryon distributions. A deep baryonic potential shortens the phase of SIDM core expansion and triggers core contraction. This effect can be further enhanced by a large self-scattering cross section. We find the final SIDM density profile is sensitive to the baryonic concentration and the strength of dark matter self-interactions. Assuming a spherical initial halo, we also study evolution of the SIDM halo shape together with the density profile. The halo shape at later epochs deviates from spherical symmetry due to the influence of the non-spherical disc potential, and its significance depends on the baryonic contribution to the total gravitational potential, relative to the dark matter one. In addition, we construct a multi-component model for the Milky Way, including an SIDM halo, a stellar disc and a bulge, and show it is consistent with observations from stellar kinematics and streams.

\end{abstract}
\begin{keywords}
methods: numerical-galaxies: evolution-galaxies: formation-galaxies: structure-cosmology: theory.
\end{keywords}
\section{Introduction}

The $\Lambda$ cold dark matter (\lcdm) model, where DM is assumed to be collisionless, is the leading theory for the growth of structure and formation of galaxies in our Universe. It fits the spectrum of matter fluctuations in the early universe with extraordinary precision \citep{planck2014} and explains many important aspects of galaxy formation
and evolution~\citep{Springel:2006vs,TrujilloGomez:2010yh,frenk2012,2014MNRAS.444.1518V,Vogelsberger:2014kha}. However, the success of \lcdm~does not preclude the possibility that DM may have strong self-interactions. When DM particles have a scattering cross section per unit mass, $\sigma_{\rm x}/m_{\rm x}\sim1~{\rm cm^2/g}$, DM collisions occur multiple times in the inner halo over the cosmological timescale and make distinct departures from the CDM predictions, while the outer halo remains collisionless, retaining the large-scale predictions of CDM~\citep{spergel1999,yoshida2000,dave2001,colin2002,Volgesberger2012,rocha2013}. In fact, SIDM has been motivated to address outstanding discrepancies between observations on galactic scales and CDM predictions, see~\cite{bullock2017} for a recent review on the CDM challenges and~\cite{2017arXiv170502358T} on their solutions within SIDM. In particular, it has been shown that the diverse shapes of  galactic rotation curves~\citep{Oman,deNaray:2009xj} can be explained naturally in the SIDM model~\citep{2017PhRvL.119k1102K,creasey2017}, while being consistent with observations in galaxy clusters~\citep{kaplinghat2015}.

DM self-interactions kinematically thermalize the inner halo and lead to distinct features in the halo properties. For dwarf galaxies, where DM dominates over all radii, SIDM thermalization leads to a large density core, and the stellar distribution is more extended in SIDM than CDM and there is tight correlation between the DM core size and the stellar one~\citep{Vogelsberger_2014}. While for baryon-dominated systems, the thermalization can significantly increase the central SIDM density and the inner halo shape follows the baryon distribution due to the influence of the baryonic potential~\citep{kaplinghat2014a}, a radical deviation from SIDM-only simulations. Moreover,~\citet{kaplinghat2014a} argued that once the inner halo reaches equilibrium, the inner SIDM profile can be modeled as an isothermal distribution that is sensitive to the final baryonic potential, but not the formation history. This has motivated a number of isolated simulations to test the response of the SIDM halo to the baryonic potential~\citep{elbert2016,creasey2017}, where they assumed a CDM halo and a stellar potential as the initial condition. Recently,~\cite{2017arXiv171109096R} performed cosmological hydrodynamical SIDM simulations of galaxy clusters and explicitly confirmed this expectation. In addition, although in the presence of strong baryonic feedback both CDM and SIDM could lead to similar density profiles~\citep{Fry:2015rta}, the internal structure of the SIDM halo is more robust to the inclusion of baryonic feedback, compared to its CDM counterpart, due to the rapid energy redistribution caused by the DM collisions~\citep{2017MNRAS.472.2945R}. 

In this paper, we utilize isolated simulations to study the response of the SIDM halo in the presence of the baryonic potential for Milky Way (MW)-sized galaxies, where the baryonic contribution to the potential is important. Our goal is to understand the interplay between the DM self-scattering strength and the baryonic concentration in shaping the SIDM distribution, and the significance of the potential in altering evolution history of the SIDM halo. In the first two sets of simulations, we vary both the baryonic concentration and $\sigma_{\rm x}/m_{\rm x}$ in the range of $0.5\textup{--}5~{\rm cm^2/g}$, and study the variation of the SIDM predictions in the density profile and shape as a function of the cross section. In the presence of baryons, the central density of an SIDM halo no longer decreases  monotonically with increasing $\sigma_{\rm x}/m_{\rm x}$, as expected in the SIDM-only case for $\sigma_{\rm x}/m_{\rm x}$ we take. Accordingly, the SIDM halo shape varies with $\sigma_{\rm x}/m_{\rm x}$ even for the same baryonic potential. Our results indicate that inferring $\sigma_{\rm x}/m_{\rm x}$ from stellar kinematics of luminous galaxies, where the baryons dominate the potential, could be challenging. 

In our third set of simulations, we construct a realistic MW mass model, including an SIDM halo, a stellar bulge and disc. We fix $\sigma_{\rm x}/m_{\rm x}=1~{\rm cm^2/g}$ and carefully adjust the model parameters to reproduce the mass model inferred from the stellar kinematics. We then make a detailed comparison between the halo shape predicted in our model and those inferred from observations.

The structure of this paper is as follows. In Sec.~\ref{sec:section2}, we discuss the numerical details of our simulations
and the methodology used to quantify the halo shapes. In Sec.~\ref{sec:halo-disc}, we use our code to explore the evolution of a MW-sized halo with a stellar disc and measure the effect of the radial length scale of the disc. In Sec.~\ref{sec:MW}, we compare our predictions for an SIDM MW halo against those from CDM simulations and those inferred from observations of stellar streams. We conclude and summarize our results in Sec.~\ref{sec:conclusion}.  
	
\section{Simulations and halo shape algorithms}
\label{sec:section2}
\subsection{Numerical Simulations}

We carry out N-body simulations using the code {\sc Arepo}~\citep{Springel2010}. 
Gravity modules in Arepo are a modified version of {\sc GADGET-2} and 
{\sc Gadget-3} \citep{springel2005}. 
We use the algorithm developed in~\citet{Volgesberger2012,2016MNRAS.460.1399V} to model DM self-interactions. This is a Monte Carlo-based method, where at each time step a particle may pairwise scatter with any of its nearest neighbors. We assume a velocity-independent constant cross section in our simulations. This is a good approximation, since the observationally self-scattering cross section varies mildly across galactic scales~\citep{kaplinghat2015} and we mainly focus on isolated simulations for a given halo mass. 
We evolve our simulations for $10\;{\rm Gyr}$, slightly shorter than Hubble time scale ($H_{0}^{-1}\approx 13.96~{\rm Gyr}$) in order to account for the assembly of the primordial galactic halo. 

Following~\citet{creasey2017}, we model the baryonic component in our simulations as a static potential. This approach ignores the back-reaction of the halo evolution on the baryons, an effect expected to be sub-dominant, since we are interested in the systems that the final baryonic distribution is known. We consider two models for the baryonic potential. One is the Miyamoto-Nagai (MN) disc~\citep{Miyamoto-Nagai},

\begin{equation}
\Phi_{\text{MN}}(R,z)\ =\ \frac{-G\ M_{\text{d}}}{\sqrt{R^2\ +\left( R_{\text{d}}
+ \sqrt{z_{\text{d}}^2+z^2}\right)^2}}, 
\label{eq:Miyamoto_nagai}
\end{equation}
where $M_{\rm d}$ is the disc mass, $R_{\rm d}$ the disc scale length and $z_{\rm d}$ the disc scale height. The implementation of the MN disc in {\sc Arepo} is as described in~\citet{creasey2017}. We also consider a Hernquist bulge potential~\citep{hernquist-1990}
\begin{equation}
\Phi_{\rm Hernquist} = -\frac{G M_{\rm H}}{r+r_{\rm H}} \, ,
\end{equation}
where $M_{\rm H}$ is the bulge mass and $r_{\rm H}$ is the scale length.

We run three sets of simulations, varying the baryonic component and the strength of the cross section. In the first two sets, we only include the MN disc ($M_{\rm d}=6.4\times10^{10}\;{M}_{\odot}$) with two disc scale lengths, $R_{\rm d}=3\;{\rm kpc}$ (compact disc) and $R_{\rm d}=6\;{\rm kpc}$ (extended disc). In both cases, we fix $z_{\rm d}=0.3 R_{\rm d}$. The DM self-scattering cross section is chosen to be $\sigma_{\rm x}/m_{\rm x}\ =0,\ 0.5,\ 1,\ 3$, and $5\; {\rm cm^2/g}$, i.e., $10$ simulations in total. For the {\it initial} halo component, we assume a spherical NFW profile~\cite{NFW}, and take the halo parameters as $r_{\rm s}=37.03~{\rm kpc}$ and $\rho_{\rm s}=2.95\times10^6M_\odot/{\rm kpc^3}$. The mass ratio of the disc to the halo is motivated by the baryonic Tully-Fisher relation~\citep{Lelli}. We use the publicly available code {\sc SpherIC}, introduced in~\cite{garrison-kimmel2013}, to generate the initial conditions. It truncates the outer halo profile exponentially  at $r_{\rm cut}$ to avoid mass divergence. We take $r_{\rm cut}\approx 250~{\rm kpc}$, close to the virial radius of our initial CDM halo. We fix the gravitational softening length to be $\epsilon= 125~\rm{pc}$ and the mass resolution $m_{\rm p}=1.32\times 10^{6} M_{\odot}$. We include $2$ million mass particles in our simulations, necessary for resolving the innermost regions, resulting in a halo mass of $2.64\times10^{12}M_{\odot}$.

For the third set, we include both an MN disc and a Hernquist bulge to model the baryon distribution in the MW. The disc parameters are $M_{\rm d}=6.98\times10^{10}\ M_{\odot}$, $R_{\rm d}=3.38~{\rm kpc}$ and $z_{\rm d}=0.2 R_{\rm d}$ for the disc. The bulge ones are $M_{\rm H}=1.05\times10^{10}\ M_{\odot}$ and $r_{\rm H}=0.46~{\rm kpc}$. The {\it initial} halo parameters are $r_{\rm s}=42.18~{\rm kpc}$ and $\rho_{\rm s}=1.39\times10^6M_\odot/{\rm kpc^3}$. We have chosen these parameters to reproduce the MW mass model presented in~\cite{mcmillan} (hereafter McM11), see Sec.~\ref{sec:MW} for details. The baryon-model parameters used in our simulations are summarized in Table~\ref{table:bar_param}. We choose $r_{\rm cut}=100$ kpc, $\epsilon=125~\rm{pc}$ and $m_{\rm p}=5.76\times 10^{5} M_{\odot}$. We simulate $2$ million mass particles and the total halo mass is $1.15\times10^{12}M_{\odot}$.

Additionally, we have run cosmological zoom-in SIDM simulations for $5$ MW-mass Aquarius halos~\citep{springel2008} with the initial conditions taken from~\citep{Volgesberger2012,zavala2013}. 
We will present these simulation results in Sec.~\ref{sec:MW} for comparison.

\begin{table}
\begin{center}
\begin{tabular}{ccc}
\hline
$\text{Component}$ & Mass ($10^{10}\ M_{\odot}$) & Length scale (kpc)\\
\hline
\hline
Extended Disc  & 6.4 & 6.0 \\
Compact Disc & 6.4 & 3.0  \\
MW-like Disc & 6.98 & 3.38  \\
MW-like Bulge & 1.05 & 0.46 \\
\hline
\end{tabular}
\end{center}
\caption{Parameters of static potentials used in the three sets of simulations.}
\label{table:bar_param}
\end{table}
\subsection{Halo shape algorithm}
\label{sub:algo}
\begin{figure}
\includegraphics[width=\columnwidth]{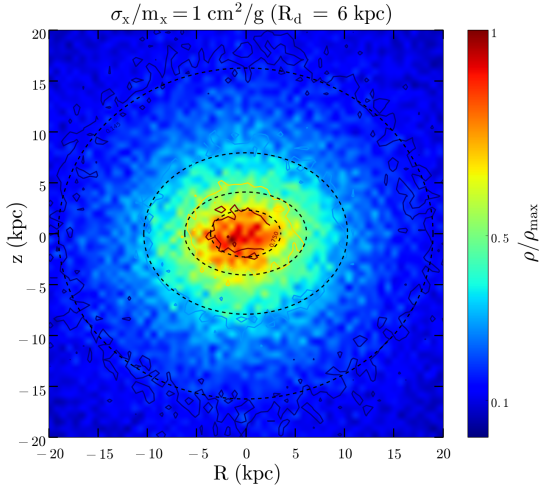}
\caption{Comparison between ellipsoids (dashed) and isodensity contours (solid) for the simulation with $\sigma_{\rm x}/m_{\rm x}=1~{\rm cm^2/g}$ and $R_{\rm d}=6~{\rm kpc}$. The horizontal axis is the major axis, while the vertical one is the minor axis aligned with the symmetry axis of the baryonic disc. The color bar shows the density scaled to $\rho_{\rm max}$, the maximal central DM density.}
\label{fig:iso_dens_vs_ellipsoids}
\end{figure}
We use the method introduced in \citet{Dubinsky_carlberg} (see also~\citealp{allgood2005}) to calculate the ellipticity of the simulated halos. It constructs the axial ratio for the best-fitting ellipsoid as a function of the major axis length. This method is an iterative one where at each iteration the reduced inertia tensor is determined for the set of 
particles within the previous ellipsoid, and then a new ellipsoid is determined from this tensor.
Specifically, if we denote the major axis length $a$, then at each iteration the reduced inertia tensor is given by 
\begin{equation}
I_{\rm ij} = \sum_{k:d_{\rm k} < a} \frac{r_{\rm k,i}\times r_{\rm k,j}}{d_{\rm k}^2}
\end{equation}
where $r_{\rm k,i}$ denotes the coordinate ${\rm i}$ of particle ${\rm k}$, and $d_{\rm k}$ is the elliptical radius found from the previous inertia tensor. We have $d_{\rm k} = \sqrt{x_{k}^2 + (y_k/q)^2 + (z_k/s)^2}$, where $x$, $y$ and $z$ are the coordinates along the major-, intermediate- and minor-axes of the ellipsoid, and $q=b/a$, $s=c/a$ are the axial ratios of the intermediate- and minor-to-major axes, respectively. After diagonalizing the inertia tensor with eigenvalues (ascending) $\{\lambda_1, \lambda_2, \lambda_3\}$, we have $q=\sqrt{\lambda_2/\lambda_3}$ and $s=\sqrt{\lambda_1/\lambda_3}$. In the initial iteration, the ellipsoid is set to a sphere, i.e. $q=s=1$. This process is continued until some convergence criteria, which we take it to be $10^{-6}$ on the difference between successive iterations, is satisfied. We note that if the number of DM particles in an ellipsoid is too small, typically  less than $1000$, the result from this method is not accurate (see Appendix~\ref{sec:A1}).

Fig.~\ref{fig:iso_dens_vs_ellipsoids} shows a comparison between isodensity contours and ellipsoids 
for an example, where $R_{\rm d} = 6$ kpc and $\sigma_{\text{x}}/m_{\text{x}}= 1\;{\rm cm^2/g}$. We see the overall agreement between the two methods is excellent.

\begin{figure*}
\includegraphics[width=\textwidth]{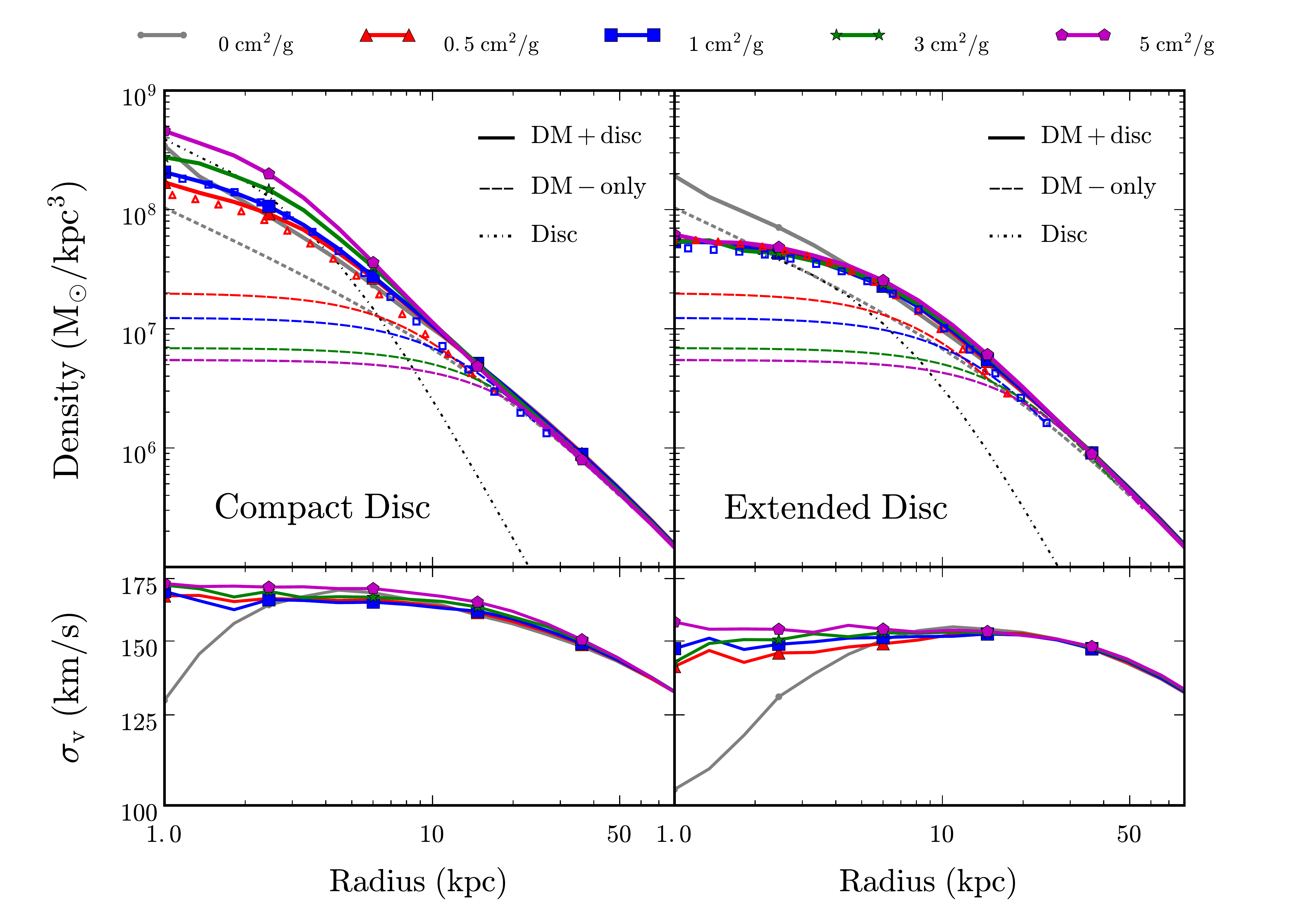}
\caption{Simulated DM density (top, solid) and velocity dispersion (bottom) profiles for $\sigma_{\rm x}/m_{\rm x} = 0$ (gray), $0.5$ (red triangles), $1$ (blue squares), $3$ (green stars), and $5~{\rm cm^2/g}$ (magenta pentagons), in the presence of compact (left) and extended (right) discs. On the top panels, we also plot the corresponding NFW initial condition (short dashed) and the SIDM density profiles derived from the analytical method without including the stellar potential (long dashed, the same color scheme as the solid ones), together with the disc mass profile (dash dotted). For comparison, we also show the SIDM density profiles derived from the analytical method with a thin-disc model for $\sigma_{\rm x}/m_{\rm x}=0.5$ and $1~{\rm cm^2/g}$ (open triangles and squares).}
\label{fig:density}
\end{figure*}
\begin{figure*}
\includegraphics[width=0.95\columnwidth]{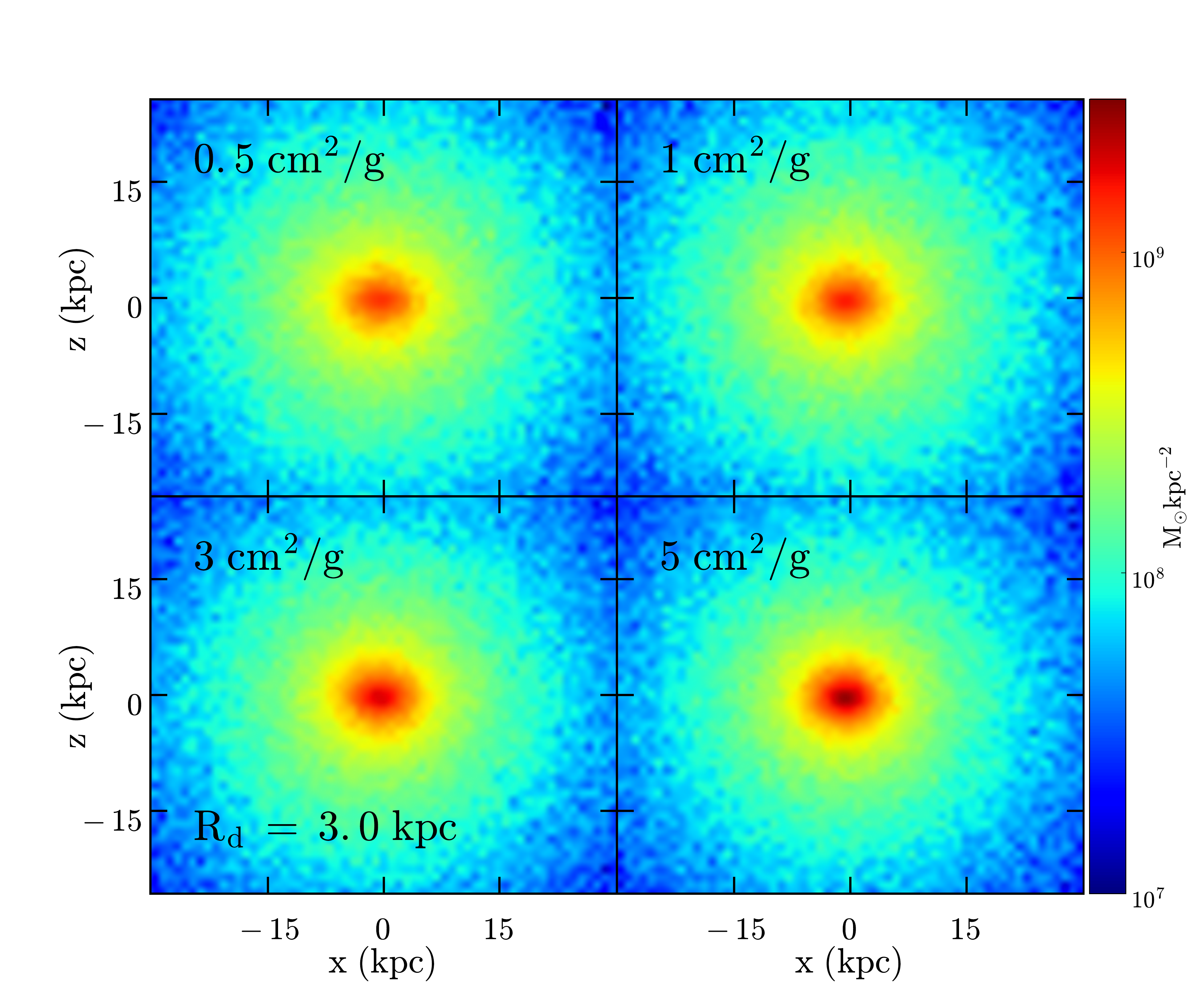}
\includegraphics[width=0.95\columnwidth]{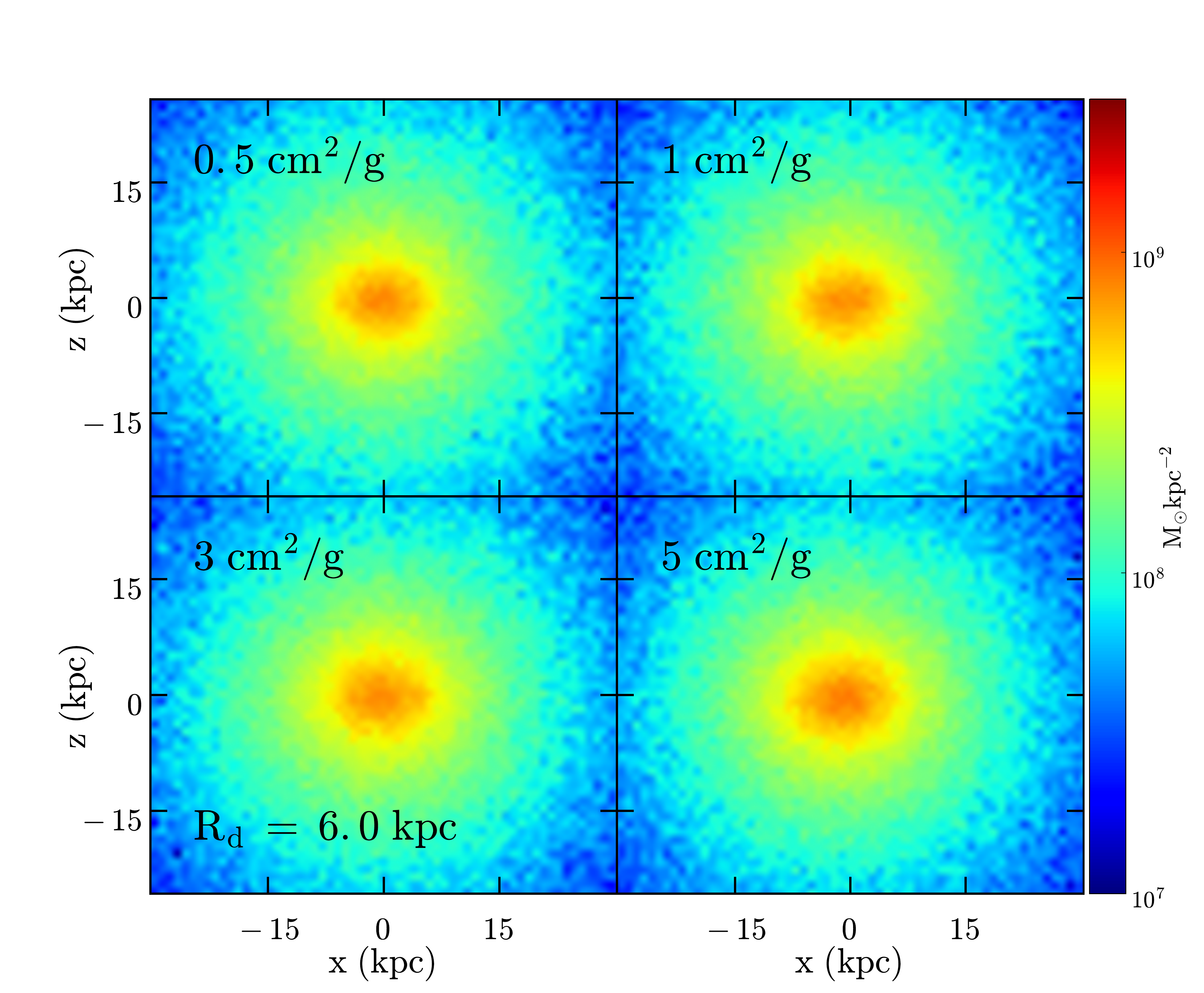}
\caption{Projected SIDM distributions in edge-on views for compact (left) and extended (right) discs with different SIDM cross sections as given in the legend.}
\label{fig:shape1}
\end{figure*}

\begin{figure*}
\includegraphics[width=\textwidth]{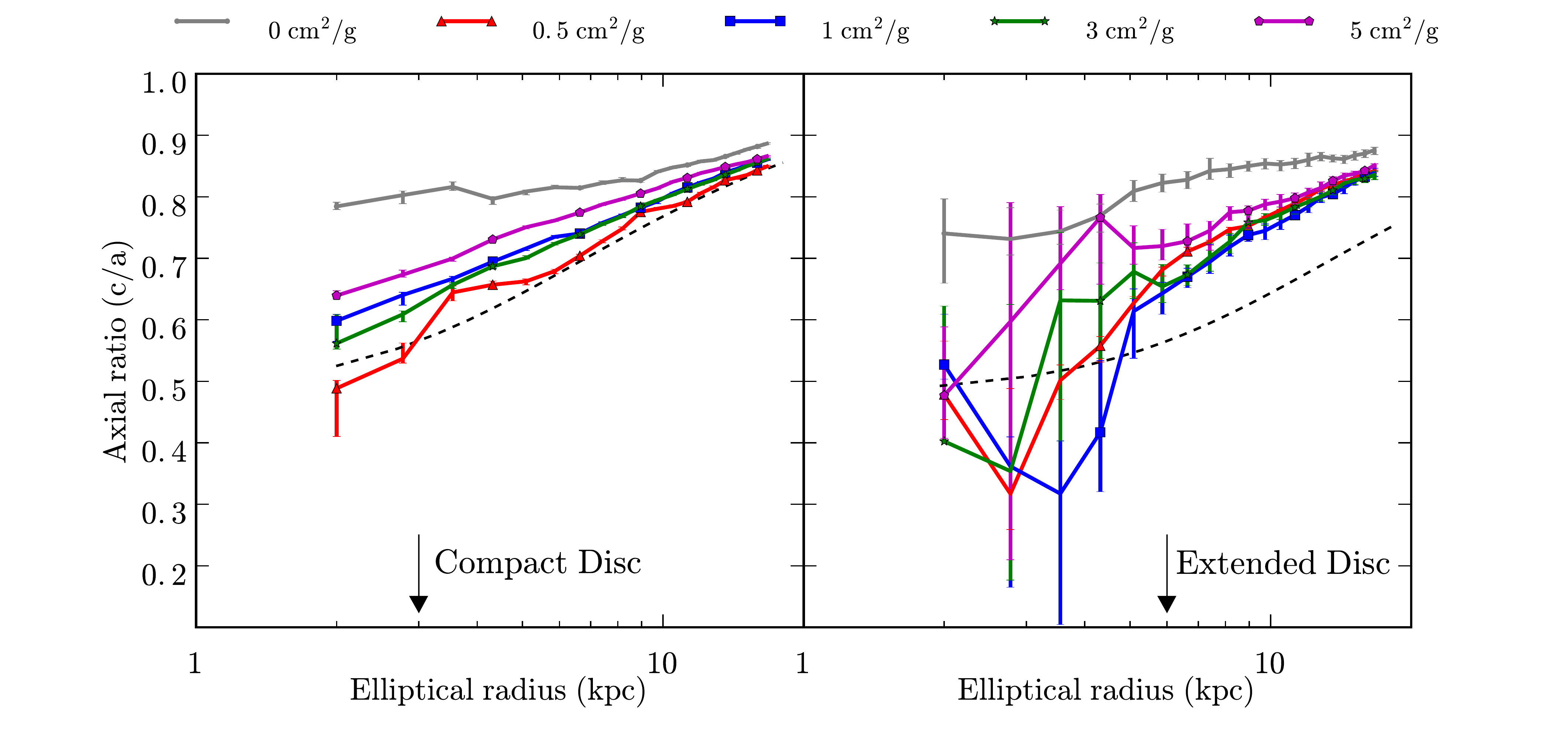}
\caption{Ratio of minor-to-major axes, $c/a$, vs. the elliptical radius for the compact (left) and extended (right) discs with different cross sections (solid). We also plot $c/a$ for the isothermal profile when the disc dominates the gravitational potential (dashed), $\rho_{\rm DM}\propto\exp[{-\Phi_{\rm MN}(R,z)/\sigma^2_{\rm v}}]$, where we take the central DM velocity dispersion for $\sigma_{\rm x}/m_{\rm x}=0.5~{\rm cm^2/g}$, i.e., $\sigma_{\rm v}\approx168$ (compact) and $150~{\rm km/s}$ (extended), as shown in Fig.~\ref{fig:density} (bottom). The arrow denotes the scale radius of the stellar disc. We calculate the numerical errors using the bootstrapping method. }
\label{fig:shape2}
\end{figure*}

\begin{figure*}
\includegraphics[width=\columnwidth]{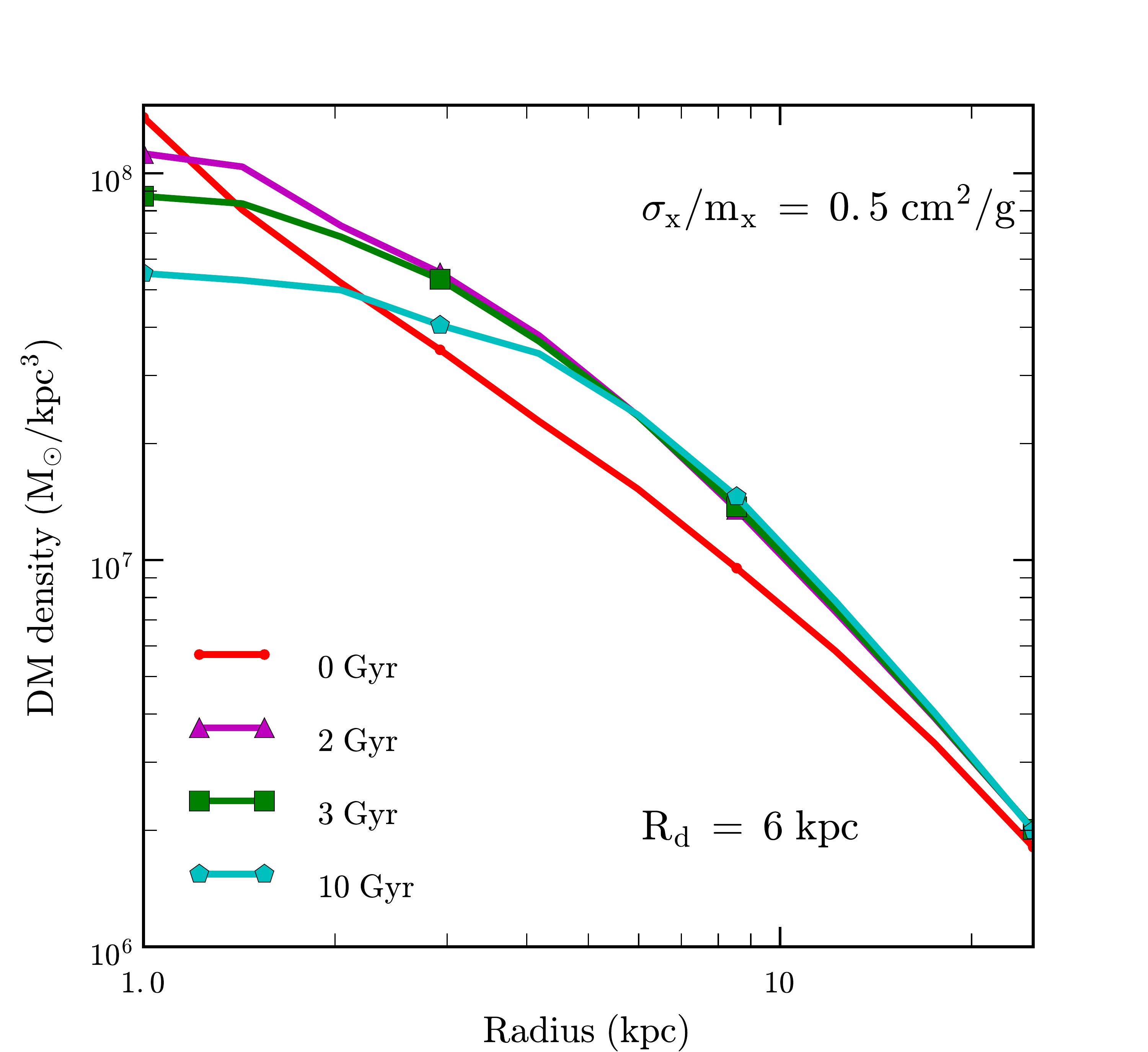}
\includegraphics[width=\columnwidth]{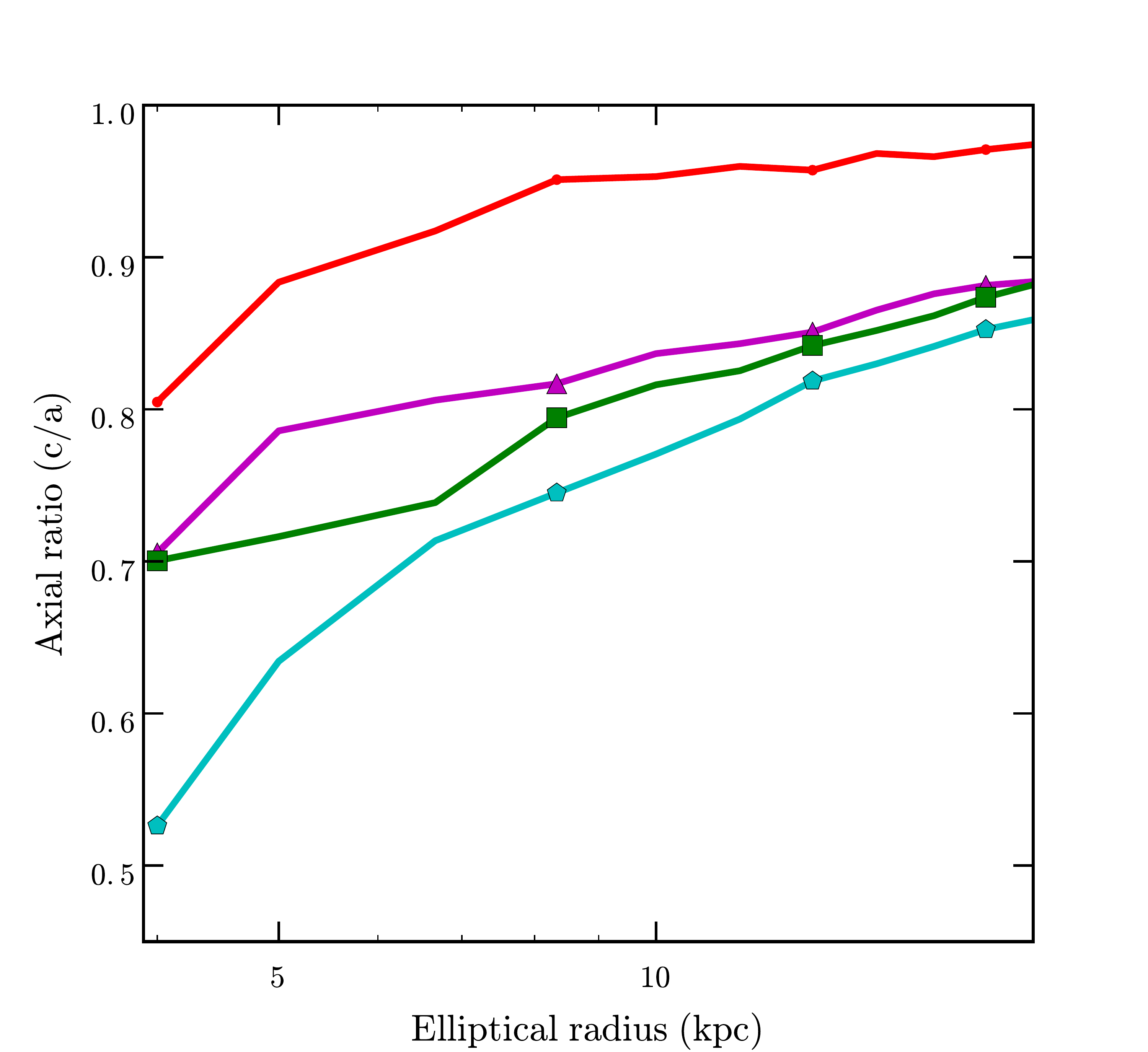}
\\
\includegraphics[width=\columnwidth]{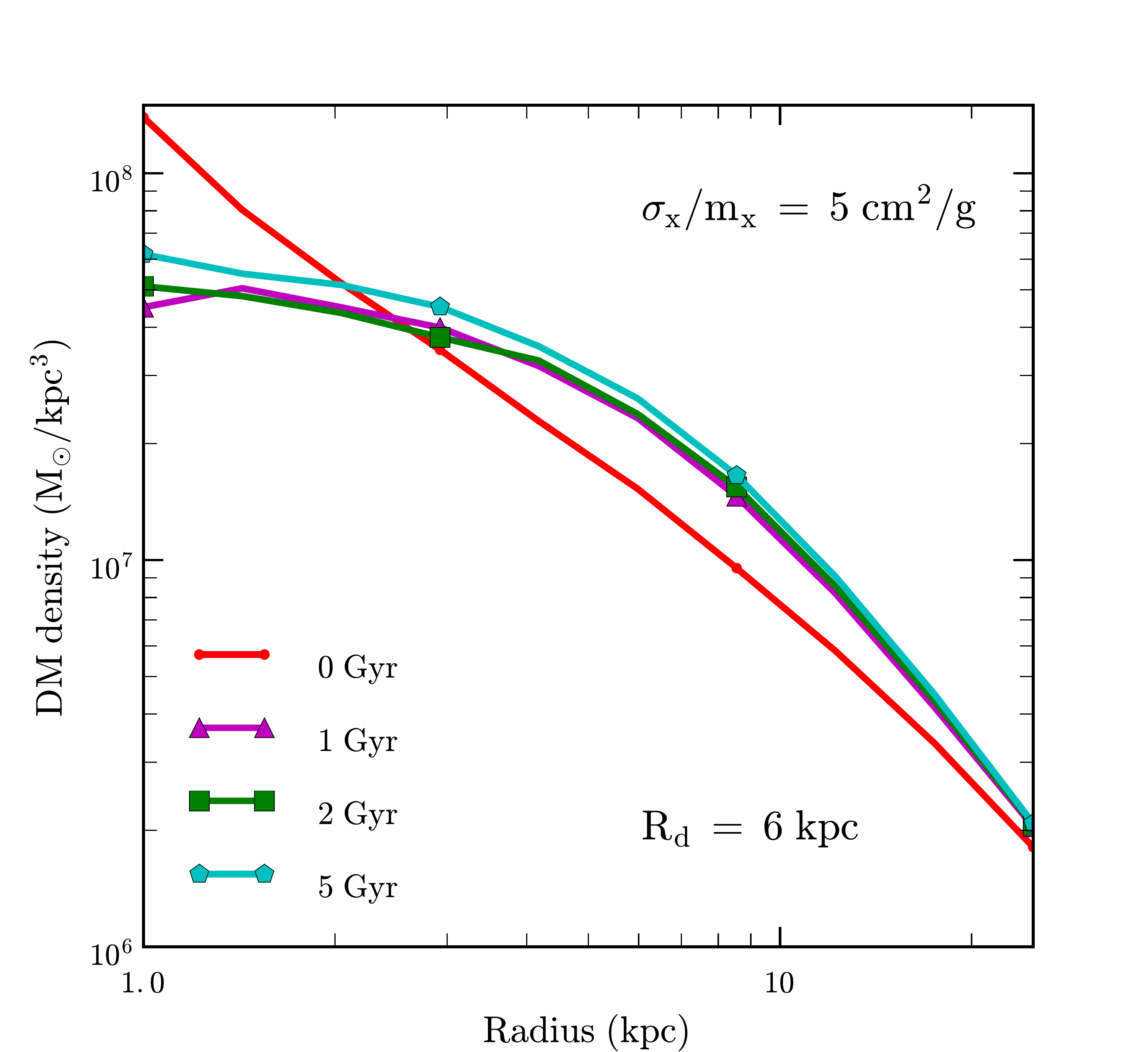}
\includegraphics[width=\columnwidth]{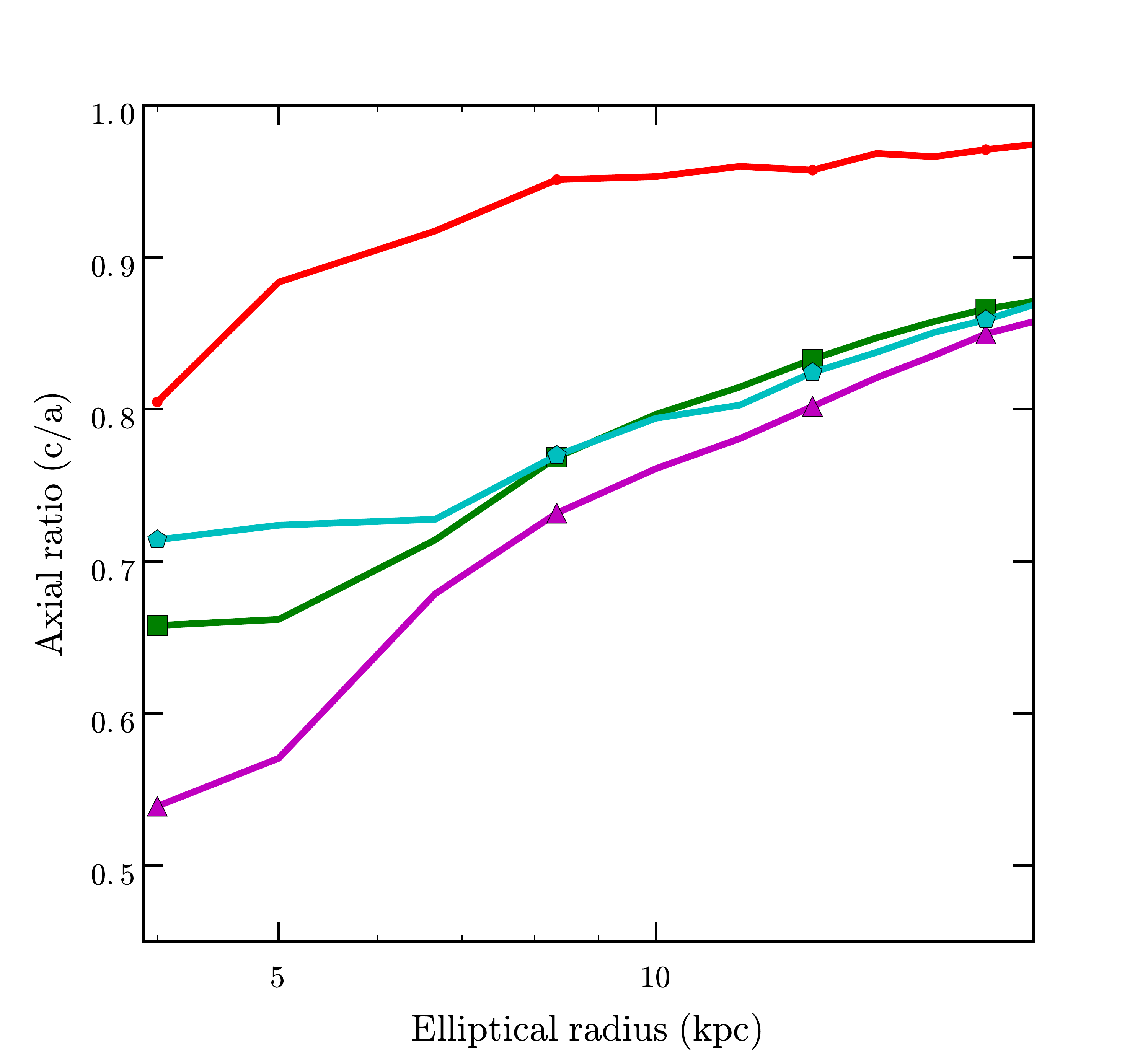}
\\
\includegraphics[width=\columnwidth]{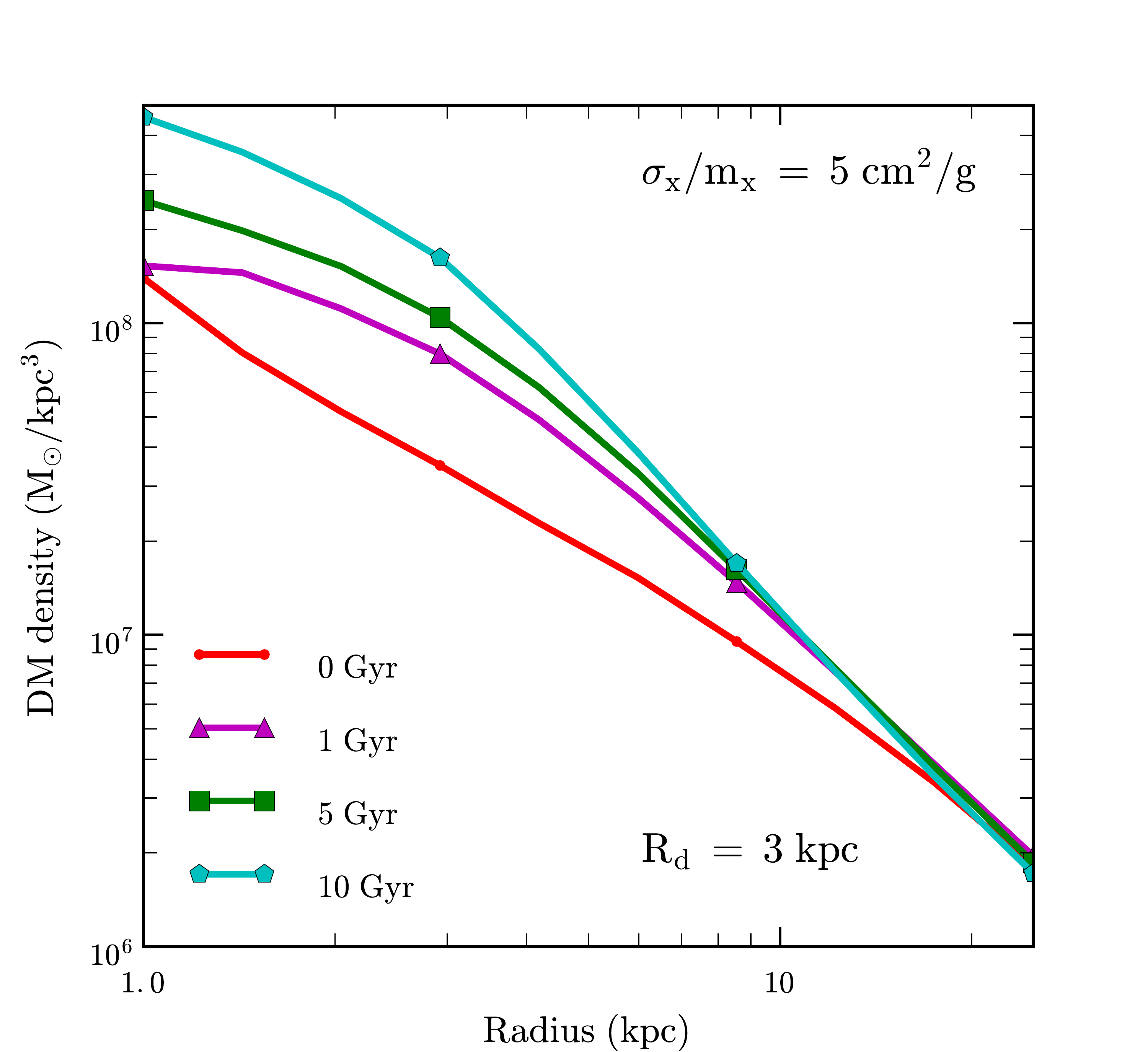}
\includegraphics[width=\columnwidth]{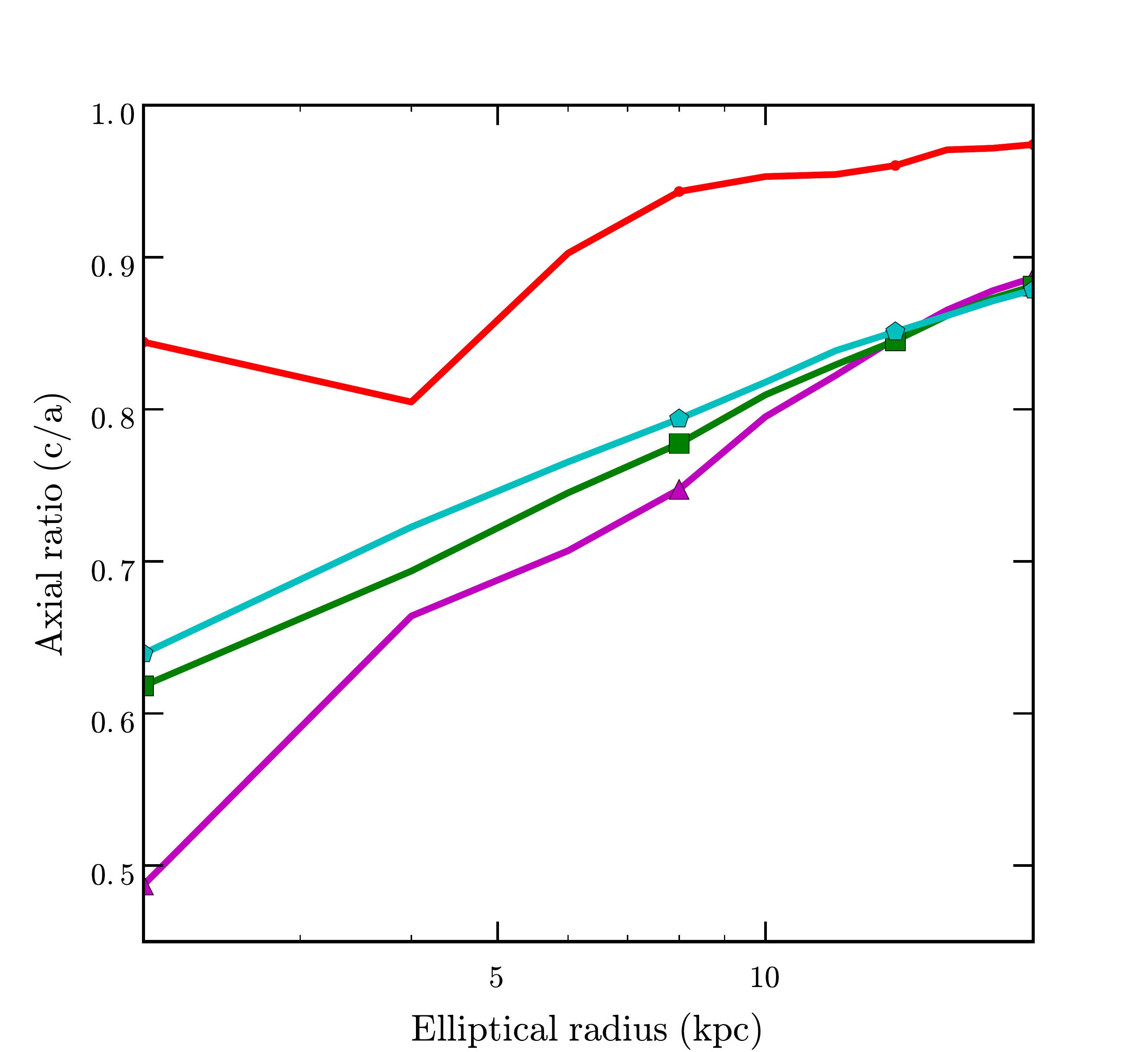}
\caption{Time evolution of the DM density (left) and the halo shape (right) profiles for three representative examples, including simulations with the extended disc for $\sigma_{\rm x}/m_{\rm x}=0.5~{\rm cm^2/g}$ (top) and $5~{\rm cm^2/g}$ (middle), and the compact one for $\sigma_{\rm x}/m_{\rm x}=5~{\rm cm^2/g}$ (bottom). Different colors and marker styles denote different evolution epochs, and both left and right panels have the same color and marker scheme.}
\label{fig:time_evol}
\end{figure*}

\section{SIDM halo properties with a stellar disc}
\label{sec:halo-disc}

\subsection{Density profiles}
\label{sub:density}

Fig.~\ref{fig:density} shows the DM density (top) and velocity dispersion (bottom) profiles for $R_{\rm d}=3~{\rm kpc}$ (left) and $6~{\rm kpc}$ (right). The solid curves are from our simulations for different values of $\sigma_{\rm x}/m_{\rm x}$ in the presence of the stellar disc. For comparison, we also plot the SIDM density profiles (dashed) without the disc potential, calculated with the analytical method in~\cite{kaplinghat2014a,kaplinghat2015}. 

In both cases, the presence of a baryonic potential increases the SIDM density profile and reduces the core size, and the effect is more significant if the baryonic concentration is higher. For $R_{\rm d}=3~{\rm kpc}$, a larger cross section leads to a higher DM density ($0.5\textup{--}5\;{\rm cm^2/g}$), opposite to the case without baryons. For the extended disc with $R_{\rm d}=6\;{\rm kpc}$, the SIDM density profiles are almost identical, even though the $\sigma_{\rm x}/m_{\rm x}$ value changes by a factor of $10$. A deep baryonic potential also increases the DM velocity dispersion in the inner halo, as shown in the bottom panels. In the case of compact disc, it is evident that all SIDM halos are close to the threshold of mild core collapse, as the velocity dispersion profiles start to develop a negative gradient from $1$ to $10~{\rm kpc}$ at the $3\textup{--}5\%$ level. The significance is continuously enhanced when $\sigma_{\rm x}/m_{\rm x}$ changes from $0.5$ to $5~{\rm cm^2/g}$. We have checked simulation results using the analytical method, where we assume a thin disc model and use the numerical templates developed in~\cite{2017PhRvL.119k1102K}. Overall, they agree well. As an example, we show the density profiles derived from the analytical method for $\sigma_{\rm x}/m_{\rm x}=0.5~{\rm cm^2/g}$ ($1~{\rm cm^2/g}$) in Fig.~\ref{fig:density} (top), and the corresponding central DM velocity dispersions are $170~{\rm km/s}$ ($170~{\rm km/s}$) and $145~{\rm km/s}$ ($154~{\rm km/s}$) for the compact and extended cases, respectively. In the analysis, we match the inner isothermal distribution to the initial NFW profile such that the density and mass are continuous within $\sim5\%$ at the radius, where scattering occurs once over $10~{\rm Gyr}$. 


The SIDM halo has a distinct evolution history. It first undergoes a core expansion phase, during which the DM collisions transport heat towards the inner region and a central density core forms. Since a self-gravitating system has a negative heat capacity, the core will eventually contract and collapse to a singular state~\citep{2002ApJ...568..475B}. In cosmological SIDM-only simulations, mild core collapse is observed within $10~{\rm Gyr}$ when $\sigma_{\rm x}/m_{\rm x}\gtrsim10~{\rm cm^2/g}$~\citep{Volgesberger2012,elbert2015}. 
We have also checked that for isolated SIDM-only ones with an NFW profile as the initial condition, the core contraction does not occur within $10~{\rm Gyr}$ for $\sigma_{\rm x}/m_{\rm x}= 0.5\textup{--}5~{\rm cm^2/g}$, consistent with the results in~\cite{koda2011}. However, the SIDM thermalization with a deep baryonic potential can speed up this process, as shown in our simulations \citep[see also][]{elbert2016}.

We see that the presence of the stellar potential breaks the monotonic relation between the value of $\sigma_{\rm x}/m_{\rm x}$ and the central SIDM density. The effect depends on the baryonic concentration and the size of the self-scattering cross section. Our results indicate that it could be challenging to extract the $\sigma_{\rm x}/m_{\rm x}$ information from stellar kinematics of galaxies dominated by baryons.

\subsection{Halo shapes}
\label{sub:shape}

In Fig.~\ref{fig:shape1}, we show the SIDM halo surface densities for $R_{\rm d}=3\;{\rm kpc}$ (left) and $R_{\rm d}=6\;{\rm kpc}$ (right). The density contrast for the compact case is higher for different cross sections, compared to the extended one, as expected from the density profiles shown in Fig.~\ref{fig:density}. It is also evident that the simulated halos are not spherically symmetric, although their initial conditions are exactly spherical.

Fig.~\ref{fig:shape2} shows the ratio of minor-to-major axes vs. elliptical radius $\sqrt{R^2 + (z/s)^2}$ for $R_{\rm d}=3\;{\rm kpc}$ and $R_{\rm d}=6\;{\rm kpc}$ with different cross sections (solid). In all cases, the $c/a$ value deviates from $1$ and decreases towards the center ($b/a$ remains close to $1$). However, the SIDM halos are more responsive to the presence of the baryonic disc than their collisionless counterpart, and their shapes are more aligned with the axisymmetric disc potential (dashed). Interestingly, for the compact case, $c/a$ increases when the cross section increases from $0.5$ to $5\;{\rm cm^2/g}$ and the inner halo becomes {\em rounder} mildly. We can see a similar trend in the case of $R_{\rm d}=6~{\rm kpc}$, although the errors in measuring $c/a$, calculated using bootstrap method, for $r\lesssim R_{\rm d}$ are large due to the lack of enough DM particles in the central region of the halos. 

The behavior in Fig.~\ref{fig:shape2} can be understood as follows. Since the DM self-interactions thermalize the inner halo, the DM density can modeled by the isothermal distribution~\citep{kaplinghat2014a}, $\rho_{\rm DM}\propto\exp{\left[-(\Phi_{\rm DM}+\Phi_{\rm MN})/\sigma^2_{\rm v}\right]}$, where $\Phi_{\rm DM}$ and $\Phi_{\rm MN}$ are the DM and disc potentials, respectively. $\Phi_{\rm MN}$ induces the deviation from spherical symmetry of the initial NFW halo, as indicated in Fig.~\ref{fig:shape2} (dashed), and the significance depends on its magnitude relative to $\Phi_{\rm DM}$ and $\sigma^2_{\rm v}$. In the compact-disc case, the central DM density increases when $\sigma_{\rm x}/m_{\rm x}$ increases from $0.5$ to $5~{\rm cm^2/g}$, as well as the DM dispersion (very mildly), as shown in Fig.~\ref{fig:density} (left). Accordingly, the baryonic potential becomes less dominant and the inner halo becomes more spherical. Note in the compact-disc case the simulated $c/a$ profile for $\sigma_{\rm x}/m_{\rm x}=0.5~{\rm cm^2/g}$ agrees well with the isothermal profile due to the baryonic potential, because of the strong dominance of the disc in the inner regions. In addition, for $\sigma_{\rm x}/m_{\rm x}=5~{\rm cm^2/g}$ (compact), both the inner DM density and velocity dispersion are higher, compared to the CDM case, but the SIDM halo is more aspherical and aligned with the disc than the CDM one. In the extended-disc case, the halo $c/a$ profiles also follow the disc one, but not as close as the compact case, since the disc does not dominate the potential at all radii, as shown in Fig.~\ref{fig:density} (right).


\subsection{Evolution history}
\label{sub:evolution}

In this section, we take a closer look at the evolution of the SIDM halo and explicitly show that the presence of the baryonic potential does speed up core contraction and shorten the expansion phase. 

Fig.~\ref{fig:time_evol} shows the density and $c/a$ profiles at different epochs for three examples: the lowest (top) and highest (middle) cross sections in the simulations with the extended disc, and the highest cross section for the compact disc case (bottom). For $R_{\rm d}=6~{\rm kpc}$ and $\sigma_{\rm x}/m_{\rm x}= 0.5\;~{\rm cm^2/g}$, the simulated halo is on the core expansion phase over the $10\;{\rm Gyr}$ span of the simulation. In this case both central DM density and the $c/a$ ratio decrease continuously. When we increase the cross section to $5\;{\rm cm^2/g}$ (middle), the duration of the core expansion phase becomes much shorter. After about $1~{\rm Gyr}$, the halo enters the core contraction phase and the central DM density increases, as well as the ratio of minor-to-major axes. A more compact stellar disc can change the halo evolution even more dramatically, as shown in the bottom panel. The simulated halo almost never gets into the expansion phase and the central density and $c/a$ in the regions increases over time monotonically. In this case, the inner SIDM halo contains even more DM mass than its CDM counterpart. 

We conclude that the evolution history of the SIDM halo is sensitive to the presence of the baryonic potential. The final halo properties, such as the density profile and the ellipticity, depend on the baryonic concentration and the strength of DM self-interactions.

\section{Implications for the shape of the Milky Way Halo}
\label{sec:MW}
The effect of baryons on the SIDM halo can potentially be tested with observations of the MW. Here, we construct a model for the MW potential consisting of an SIDM halo, a baryonic bulge and disc. The DM halo is chosen initially as a spherical NFW profile with $r_{\rm s}=42.18~{\rm kpc}$ and $\rho_{\rm s}=1.39\times10^6M_\odot/{\rm kpc^3}$. We model the disc following an MN potential as in Eq.~\ref{eq:Miyamoto_nagai}, with disc length scale and mass specified in Table~\ref{table:bar_param} and bulge following a spherical Hernquist profile. We take the cross section as $\sigma_{\rm x}/m_{\rm x}=1\;{\rm cm^2/g}$.

\begin{figure}
\includegraphics[width=\columnwidth]{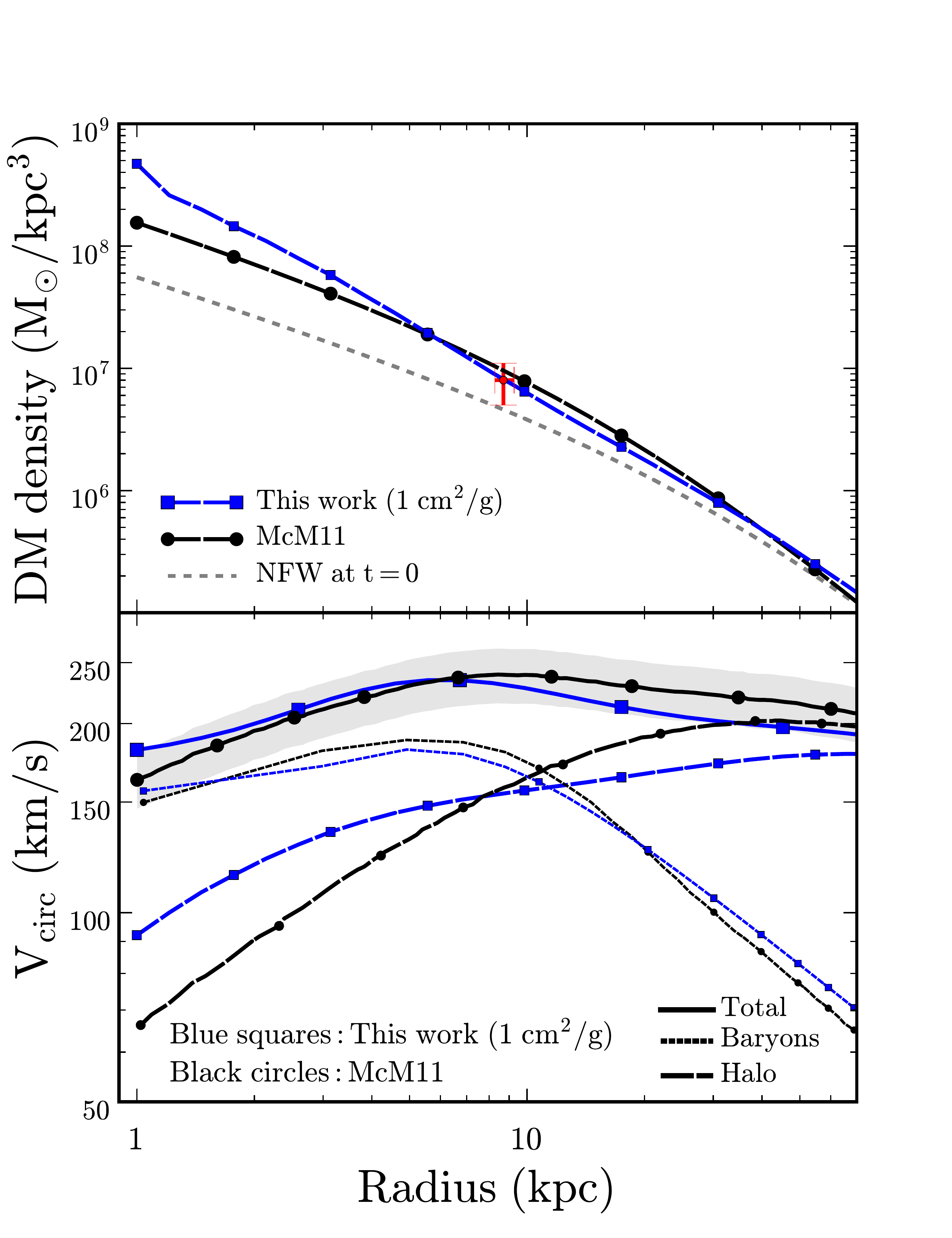}
\caption{Top: DM density profiles of simulated SIDM halo with baryons (blue squares), best-fitting halo model in~\citetalias{mcmillan} (black circles), and the NFW initial condition (gray dashed). The data point with error bars indicates the local DM density near the solar position from~\citet{Bovy2012}. Bottom: Total circular velocity profiles of our MW mass model (blue solid) and best-fitting model in~\citetalias{mcmillan} (black solid) with $10\%$ uncertainties (shaded band). Plotted also DM halo (dashed) and baryon (thin dashed) contributions.}
\label{fig:lowconcen}
\end{figure}

Top panel of Fig.~\ref{fig:lowconcen} shows the DM density profiles for our SIDM halo (blue squares), the initial NFW model (gray dashed), and the halo model in~\citetalias{mcmillan} (black circles). Our MW model reproduces well the estimates for the local DM density near the solar neighborhood from~\citet{Bovy2012} (red). Note the initial halo concentration is about $1.5~\sigma$ lower than the average for the MW mass object according to~\citet{Dutton2014} and also lower than that in~\citetalias{mcmillan}. This is a necessary choice to be consistent with observations, since SIDM thermalization significantly increases the DM density in the inner regions due to the presence of the baryonic potential. Although the inner density profile of the SIDM halo deviates from the~\citetalias{mcmillan} one, our MW mass model produces a circular velocity profile, consistent with the one in ~\citetalias{mcmillan} within $10~\%$ uncertainties, as shown in Fig.~\ref{fig:lowconcen} (bottom). 

To quantify the baryonic influence, we estimate the logarithmic slope of the SIDM density profile as $\alpha\sim-1.8$ for the range of $r=1\textup{--}2\% r_{\rm vir}$. Interestingly, this is consistent with the prediction in NIHAO hydrodynamical CDM simulations within the $1\sigma$ range~\citep{tollet2016}. Thus, for galaxies like the MW, where the baryons dominate the inner regions, both SIDM and CDM can lead to similar predictions. The result shown in Fig.~\ref{fig:lowconcen} (top) is based on $\sigma_{\rm x}/m_{\rm x}=1~{\rm cm^2/g}$. Increasing the cross section will speed up the transition from the core expansion to contraction phases, resulting a denser inner halo, as illustrated in Fig.~\ref{fig:time_evol}.

\begin{figure}
\includegraphics[width=\columnwidth]{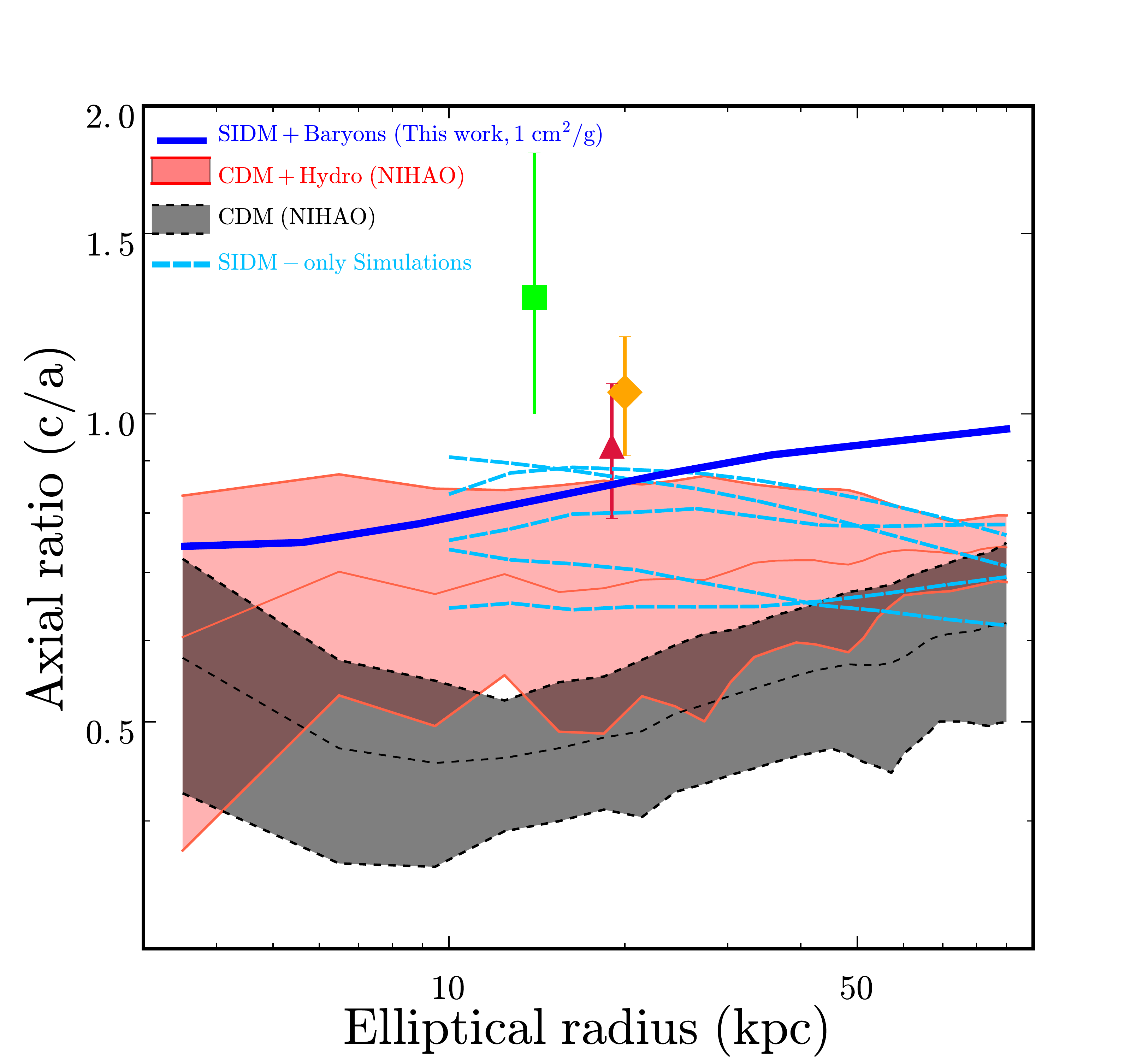}
\caption{Halo shape measurement in different numerical simulations: our MW SIDM halo model (blue solid), cosmological CDM-only (black dashed) and hydrodynamical simulations (orange solid) from the NIHAO project~\citep{butsky2016} (the shaded area represents the $1\sigma$ scatter for the halos with mass $\sim10^{12}~M_{\odot}$), and zoom-in cosmological SIDM-only simulations of $5$ Aquarius halos (long dashed). Data points with error bars are the measurements of the MW halo shape using the stellar streams, GD-1 ($c/a=1.3^{+0.5}_{-0.3}$ at $r\approx14~{\rm kpc}$, square), Pal 5 ($c/a=0.93\pm0.16$ at $r\approx19~{\rm kpc}$, pentagon), and the combined analysis of the two ($c/a=1.05\pm0.14$ for $r\lesssim20~{\rm kpc}$, diamond), taken from~\citet{bovy2016}. }
\label{fig:MW_axial_ratio}
\end{figure}

Fig.~\ref{fig:MW_axial_ratio} shows the ratio of minor-to-major axes as a function of the elliptical radius, predicted in our SIDM MW halo model with $\sigma_{\rm x}/m_{\rm x}=1~{\rm cm}^2/{\rm g}$ (blue solid). Since we assume a spherical NFW initial halo, the deviation from $c/a=1$ is caused by the disc potential. We see the disc induces a mild asphericity in the inner regions of the halo, $c/a\sim0.7\textup{--}0.8$, in good agreement with the result based on the analytical model in~\cite{kaplinghat2014a}, where the spherical boundary condition is imposed at $10~{\rm kpc}$. Our simulations also show the effect is quite extended, with $c/a$ converging back to unity only at the distance $\sim 50~{\rm kpc}$. We also present cosmological SIDM-only simulations of $5$ Aquarius halos with $\sigma_{\rm x}/m_{\rm x}=1~{\rm cm^2/g}$ (blue long dashed). Our results are in agreement with the previous ones~\citep{A.Peter,2018MNRAS.474..746B}. For the roundest halos (Aq-A and Aq-C), $c/a\gtrsim0.85$ for $r\lesssim30~{\rm kpc}$, lending support to our assumption of an initially spherical NFW halo. Since the DM self-scattering rate increases by a factor of $100$ from $30~{\rm kpc}$ to inner few ${\rm kpc}$, as indicated by the DM density profile shown in Fig.~\ref{fig:lowconcen} (top), we expect the spherical assumption is well-justified in the inner halo.

The MW halo shape has been inferred from observations of stellar streams such as GD-1 and Pal 5~\citep[e.g.,][]{koposov2010,bowden2015,pearson2015, kupper2015,bovy2016}. In Fig.~\ref{fig:MW_axial_ratio}, we show the results presented in~\citet{bovy2016} for GD-1, Pal 5 and the combined one, where an axisymmetric NFW density profile with $b/a=1$ was used to model the DM distribution in the MW. We see that the phase-space tracks of the streams are consistent with a spherical DM halo in the MW at intermediate radii, $r\sim20~{\rm kpc}$, which indicates any asphericity either intrinsic to the DM distribution or induced by the disc should be at most weak on that scale in order to accommodate the measurements. Our model predicts $c/a\sim0.85$ at $r\approx20~{\rm kpc}$, consistent with the combined constraint on $c/a$ within $\sim1.5\sigma$. 
Note that our scale height in the MN disc is $z_{\rm d}\approx0.68~{\rm kpc}$, higher than the best-fit value $\sim0.3~{\rm kpc}$ in~\citet{bovy2016}. We have estimated that taking their $z_{\rm d}$ value would reduce $c/a$ by $5\%$ at most in the inner regions and the difference becomes negligible around $10~{\rm kpc}$. 
In addition, in our MW model, the total halo mass within $20~{\rm kpc}$ is $1.28\times10^{11}M_\odot$, consistent with $M_{\rm halo}(r<20~{\rm kpc})=1.1\pm0.1\times10^{11}M_{\odot}$ measured in~\citet{bovy2016}, although our initial NFW halo has lower concentration, compared to theirs ($r_s=18.0\pm7.5~{\rm kpc}$). It would be interesting to analyze the stream data with the SIDM halo model.

For comparison, we also plot cosmological CDM-only (gray) and hydrodynamical (red) simulations from the NIHAO project~\citep{butsky2016}. Overall, the shape of individual NIHAO halos follows the median trend shown in Fig.~\ref{fig:MW_axial_ratio}. As is well-known, CDM halos from cosmological simulations are strongly triaxial~\citep{frenk1988,suto2002,hayashi2007,kuhlen2007,Vera-Ciro1,diemand2011}. Taken at face value, CDM predictions (gray) could  look at odds with the observations. However, baryons can make the CDM halo shapes more spherical (red)~\citep[see also,][]{Dubinski1994,debattista2008,kazantzidis2010,abadi2010,tissera2010}, an effect partially attributed to the change of orbits from boxy to tube or rounder loop as a result of the central concentration of baryons~\citep{debattista2008}. In the NIHAO simulations, the mean value of $c/a$ is $0.7$ for the CDM halos after including baryons, and it can reaches $0.8$ at the $1\sigma$ level of the scatter, consistent with the observations reasonably well.

Although the sphericity created by the baryons helps CDM to accommodate more easily the observational constraints, the trend with radius could be different in the two models. It seems that CDM halos plus baryons become more spherical at all radii, whereas the effects explored here in SIDM plus baryons would anticipate a flattening of the shapes towards the inner regions that follows that of the disc. Such premise, of course, ignores any effect of feedback or cosmological assembly, which may cause deviations of the system from equilibrium. Therefore, the exciting premise of using halo shape profiles to differentiate DM candidates awaits confirmation from cosmological hydrodynamical SIDM simulations. We hope such experiments will become available in the near future.

\section{Conclusions}\label{sec:conclusion}
We use isolated N-body simulations of DM halos with static disc potentials to explore the gravitational effect of
baryons on SIDM halos. We model the disc as a Miyamoto-Nagai potential embedded within an initially NFW halo with mass $\sim 10^{12}~M_\odot$. We consider different self-scattering cross sections, $\sigma_{\rm x}/m_{\rm x}=0.5,~1,~3$, and $5~{\rm cm^2/g}$ besides the special case, $\sigma_{\rm x}/m_{\rm x}=0~{\rm cm^2/g}$. In addition, we vary the radial length scale of the disc, $R_{\rm d}$, to study in detail how the DM halo responds to the baryons as a function of how relevant their contribution is to the total potential at a given radius.

In the absence of baryons, SIDM halos develop a central flat core with its density and size that depend mostly on the self-scattering cross section. We confirm that the inclusion of a disc potential can change this behavior due to SIDM thermalization with the potential, resulting in a higher core density and a smaller core than expected without the disc, a crucial effect in solving the diversity problem in SIDM~\citep{2017PhRvL.119k1102K,creasey2017},

We highlight two phases of evolution during our numerical experiments: a first stage of {\it core expansion}, during which the density core gets established due to the turn-on of the self-interactions, and a second stage of {\it core contraction} due to the gravitational effects of the baryons. The timescale for these two phases of evolution is a function of both, the cross section and the relative importance of the baryons inside the core. Higher cross sections and more compact baryonic discs (encoded in a smaller length scale) speed up the transition between the
two phases and make the timescale of core expansion shorter.

We have also studied the role of the disc potential in shaping the SIDM halo. To explore this subtle effect, we assumed an exact spherical initial NFW profile such that any departure from sphericity is due to the influence of the baryonic potential. Compared to the case of $\sigma_{\rm x}/m_{\rm x}=0~{\rm cm^2/g}$, the SIDM halos are more responsive to the potential due to the thermalization, and their final flattening is more aligned with the orientation of the disc, consistent with the expectation from the analytical method. Our simulations clearly demonstrate that the induced asphericity is mainly sensitive to the contribution of the disc to the total potential, relative to the DM one. We further confirmed this by checking the evolution history of the SIDM halos. The flattening effect is maximized during the epoch when the core has the lowest density, which coincides with the time when the disc contribution to the total potential is also maximized.

We have constructed a mass model for the MW and explored the shape prediction with observations. The model consists of a stellar disc and a bulge, embedded within a spherical SIDM halo. It reproduces observed stellar kinematics of the Galaxy within the uncertainties and the local DM density reported in the solar neighborhood. We find that the baryons are able to induce a mild flattening ($c/a \sim 0.7\textup{--}0.8$) in the inner regions but the effect weakens at larger radii. At $r \sim 20$ kpc where observational constraints seem to suggest an almost spherical halo, the effects of the disc are not strong, in agreement with the observations. We propose that the quasi sphericity of the halo at large distance is easier to accommodate in SIDM models than within the strongly triaxial structures predicted by CDM, although considering
the effects of baryons might help to reconcile CDM models with observed spherical halos. Furthermore, we argue that a study of halo shapes as a function of radius might be able to help distinguish the nature of DM, although a more stringent comparison to cosmological simulations are needed to confirm this last point. On the observational side, future surveys capable of inferring the shape of the Galactic halo within the inner $20~{\rm kpc}$ regions are promising avenues to make progress on establishing the non-canonical nature of DM.

\section*{Acknowledgments}
OS acknowledges support by NASA MUREP Institutional Research Opportunity (MIRO) grant number NNX15AP99A and HST grant HST-AR-14582. 
HBY acknowledges support from U. S. Department of Energy  under  Grant  No. de-sc0008541  and  the  Hellman Fellows Fund. 
LVS is grateful for support from the Hellman Fellows Foundation and HST grant 
HST-AR-14582.
MV acknowledges support through an MIT RSC award, the support of the Alfred P. Sloan Foundation, and support by
NASA ATP grant NNX17AG29G.
JZ acknowledges support by a Grant of Excellence from the Iceland 
Research Fund (grant number 173929-051). 
\bibliographystyle{mnras}
\bibliography{paper}

\begin{thebibliography}{}
\makeatletter
\relax
\def\mn@urlcharsother{\let\do\@makeother \do\$\do\&\do\#\do\^\do\_\do\%\do\~}
\def\mn@doi{\begingroup\mn@urlcharsother \@ifnextchar [ {\mn@doi@}
  {\mn@doi@[]}}
\def\mn@doi@[#1]#2{\def\@tempa{#1}\ifx\@tempa\@empty \href
  {http://dx.doi.org/#2} {doi:#2}\else \href {http://dx.doi.org/#2} {#1}\fi
  \endgroup}
\def\mn@eprint#1#2{\mn@eprint@#1:#2::\@nil}
\def\mn@eprint@arXiv#1{\href {http://arxiv.org/abs/#1} {{\tt arXiv:#1}}}
\def\mn@eprint@dblp#1{\href {http://dblp.uni-trier.de/rec/bibtex/#1.xml}
  {dblp:#1}}
\def\mn@eprint@#1:#2:#3:#4\@nil{\def\@tempa {#1}\def\@tempb {#2}\def\@tempc
  {#3}\ifx \@tempc \@empty \let \@tempc \@tempb \let \@tempb \@tempa \fi \ifx
  \@tempb \@empty \def\@tempb {arXiv}\fi \@ifundefined
  {mn@eprint@\@tempb}{\@tempb:\@tempc}{\expandafter \expandafter \csname
  mn@eprint@\@tempb\endcsname \expandafter{\@tempc}}}

\bibitem[\protect\citeauthoryear{{Abadi}, {Navarro}, {Fardal}, {Babul}  \&
  {Steinmetz}}{{Abadi} et~al.}{2010}]{abadi2010}
{Abadi} M.~G.,  {Navarro} J.~F.,  {Fardal} M.,  {Babul} A.,   {Steinmetz} M.,
  2010, \mn@doi [\mnras] {10.1111/j.1365-2966.2010.16912.x}, \href
  {http://adsabs.harvard.edu/abs/2010MNRAS.407..435A} {407, 435}

\bibitem[\protect\citeauthoryear{{Allgood}, {Flores}, {Primack}, {Kravtsov},
  {Wechsler}, {Faltenbacher}  \& {Bullock}}{{Allgood}
  et~al.}{2006}]{allgood2005}
{Allgood} B.,  {Flores} R.~A.,  {Primack} J.~R.,  {Kravtsov} A.~V.,  {Wechsler}
  R.~H.,  {Faltenbacher} A.,   {Bullock} J.~S.,  2006, \mn@doi [\mnras]
  {10.1111/j.1365-2966.2006.10094.x}, \href
  {http://adsabs.harvard.edu/abs/2006MNRAS.367.1781A} {367, 1781}

\bibitem[\protect\citeauthoryear{{Balberg}, {Shapiro}  \& {Inagaki}}{{Balberg}
  et~al.}{2002}]{2002ApJ...568..475B}
{Balberg} S.,  {Shapiro} S.~L.,   {Inagaki} S.,  2002, \mn@doi [\apj]
  {10.1086/339038}, \href {http://adsabs.harvard.edu/abs/2002ApJ...568..475B}
  {568, 475}

\bibitem[\protect\citeauthoryear{{Bovy} \& {Tremaine}}{{Bovy} \&
  {Tremaine}}{2012}]{Bovy2012}
{Bovy} J.,  {Tremaine} S.,  2012, \mn@doi [\apj] {10.1088/0004-637X/756/1/89},
  \href {http://adsabs.harvard.edu/abs/2012ApJ...756...89B} {756, 89}

\bibitem[\protect\citeauthoryear{{Bovy}, {Bahmanyar}, {Fritz}  \&
  {Kallivayalil}}{{Bovy} et~al.}{2016}]{bovy2016}
{Bovy} J.,  {Bahmanyar} A.,  {Fritz} T.~K.,   {Kallivayalil} N.,  2016, \mn@doi
  [\apj] {10.3847/1538-4357/833/1/31}, \href
  {http://adsabs.harvard.edu/abs/2016ApJ...833...31B} {833, 31}

\bibitem[\protect\citeauthoryear{{Bowden}, {Belokurov}  \& {Evans}}{{Bowden}
  et~al.}{2015}]{bowden2015}
{Bowden} A.,  {Belokurov} V.,   {Evans} N.~W.,  2015, \mn@doi [\mnras]
  {10.1093/mnras/stv285}, \href
  {http://adsabs.harvard.edu/abs/2015MNRAS.449.1391B} {449, 1391}

\bibitem[\protect\citeauthoryear{{Brinckmann}, {Zavala}, {Rapetti}, {Hansen}
  \& {Vogelsberger}}{{Brinckmann} et~al.}{2018}]{2018MNRAS.474..746B}
{Brinckmann} T.,  {Zavala} J.,  {Rapetti} D.,  {Hansen} S.~H.,   {Vogelsberger}
  M.,  2018, \mn@doi [\mnras] {10.1093/mnras/stx2782}, \href
  {http://adsabs.harvard.edu/abs/2018MNRAS.474..746B} {474, 746}

\bibitem[\protect\citeauthoryear{{Bullock} \& {Boylan-Kolchin}}{{Bullock} \&
  {Boylan-Kolchin}}{2017}]{bullock2017}
{Bullock} J.~S.,  {Boylan-Kolchin} M.,  2017, \mn@doi [\araa]
  {10.1146/annurev-astro-091916-055313}, \href
  {http://adsabs.harvard.edu/abs/2017ARA%26A..55..343B} {55, 343}

\bibitem[\protect\citeauthoryear{{Butsky} et~al.,}{{Butsky}
  et~al.}{2016}]{butsky2016}
{Butsky} I.,  et~al., 2016, \mn@doi [\mnras] {10.1093/mnras/stw1688}, \href
  {http://adsabs.harvard.edu/abs/2016MNRAS.462..663B} {462, 663}

\bibitem[\protect\citeauthoryear{{Col{\'{\i}}n}, {Avila-Reese}, {Valenzuela}
  \& {Firmani}}{{Col{\'{\i}}n} et~al.}{2002}]{colin2002}
{Col{\'{\i}}n} P.,  {Avila-Reese} V.,  {Valenzuela} O.,   {Firmani} C.,  2002,
  \mn@doi [\apj] {10.1086/344259}, \href
  {http://adsabs.harvard.edu/abs/2002ApJ...581..777C} {581, 777}

\bibitem[\protect\citeauthoryear{{Creasey}, {Sameie}, {Sales}, {Yu},
  {Vogelsberger}  \& {Zavala}}{{Creasey} et~al.}{2017}]{creasey2017}
{Creasey} P.,  {Sameie} O.,  {Sales} L.~V.,  {Yu} H.-B.,  {Vogelsberger} M.,
  {Zavala} J.,  2017, \mn@doi [\mnras] {10.1093/mnras/stx522}, \href
  {http://adsabs.harvard.edu/abs/2017MNRAS.468.2283C} {468, 2283}

\bibitem[\protect\citeauthoryear{{Dav{\'e}}, {Spergel}, {Steinhardt}  \&
  {Wandelt}}{{Dav{\'e}} et~al.}{2001}]{dave2001}
{Dav{\'e}} R.,  {Spergel} D.~N.,  {Steinhardt} P.~J.,   {Wandelt} B.~D.,  2001,
  \mn@doi [\apj] {10.1086/318417}, \href
  {http://adsabs.harvard.edu/abs/2001ApJ...547..574D} {547, 574}

\bibitem[\protect\citeauthoryear{{Debattista}, {Moore}, {Quinn}, {Kazantzidis},
  {Maas}, {Mayer}, {Read}  \& {Stadel}}{{Debattista}
  et~al.}{2008}]{debattista2008}
{Debattista} V.~P.,  {Moore} B.,  {Quinn} T.,  {Kazantzidis} S.,  {Maas} R.,
  {Mayer} L.,  {Read} J.,   {Stadel} J.,  2008, \mn@doi [\apj]
  {10.1086/587977}, \href {http://adsabs.harvard.edu/abs/2008ApJ...681.1076D}
  {681, 1076}

\bibitem[\protect\citeauthoryear{{Diemand} \& {Moore}}{{Diemand} \&
  {Moore}}{2011}]{diemand2011}
{Diemand} J.,  {Moore} B.,  2011, \mn@doi [Advanced Science Letters]
  {10.1166/asl.2011.1211}, \href
  {http://adsabs.harvard.edu/abs/2011ASL.....4..297D} {4, 297}

\bibitem[\protect\citeauthoryear{{Dubinski}}{{Dubinski}}{1994}]{Dubinski1994}
{Dubinski} J.,  1994, \mn@doi [\apj] {10.1086/174512}, \href
  {http://adsabs.harvard.edu/abs/1994ApJ...431..617D} {431, 617}

\bibitem[\protect\citeauthoryear{{Dubinski} \& {Carlberg}}{{Dubinski} \&
  {Carlberg}}{1991}]{Dubinsky_carlberg}
{Dubinski} J.,  {Carlberg} R.~G.,  1991, \mn@doi [\apj] {10.1086/170451}, \href
  {http://adsabs.harvard.edu/abs/1991ApJ...378..496D} {378, 496}

\bibitem[\protect\citeauthoryear{{Dutton} \& {Macci{\`o}}}{{Dutton} \&
  {Macci{\`o}}}{2014}]{Dutton2014}
{Dutton} A.~A.,  {Macci{\`o}} A.~V.,  2014, \mn@doi [\mnras]
  {10.1093/mnras/stu742}, \href
  {http://adsabs.harvard.edu/abs/2014MNRAS.441.3359D} {441, 3359}

\bibitem[\protect\citeauthoryear{{Elbert}, {Bullock}, {Garrison-Kimmel},
  {Rocha}, {O{\~n}orbe}  \& {Peter}}{{Elbert} et~al.}{2015}]{elbert2015}
{Elbert} O.~D.,  {Bullock} J.~S.,  {Garrison-Kimmel} S.,  {Rocha} M.,
  {O{\~n}orbe} J.,   {Peter} A.~H.~G.,  2015, \mn@doi [\mnras]
  {10.1093/mnras/stv1470}, \href
  {http://adsabs.harvard.edu/abs/2015MNRAS.453...29E} {453, 29}

\bibitem[\protect\citeauthoryear{{Elbert}, {Bullock}, {Kaplinghat},
  {Garrison-Kimmel}, {Graus}  \& {Rocha}}{{Elbert} et~al.}{2016}]{elbert2016}
{Elbert} O.~D.,  {Bullock} J.~S.,  {Kaplinghat} M.,  {Garrison-Kimmel} S.,
  {Graus} A.~S.,   {Rocha} M.,  2016, preprint, \href
  {http://adsabs.harvard.edu/abs/2016arXiv160908626E} {} (\mn@eprint {arXiv}
  {1609.08626})

\bibitem[\protect\citeauthoryear{{Frenk} \& {White}}{{Frenk} \&
  {White}}{2012}]{frenk2012}
{Frenk} C.~S.,  {White} S.~D.~M.,  2012, \mn@doi [Annalen der Physik]
  {10.1002/andp.201200212}, \href
  {http://adsabs.harvard.edu/abs/2012AnP...524..507F} {524, 507}

\bibitem[\protect\citeauthoryear{{Frenk}, {White}, {Davis}  \&
  {Efstathiou}}{{Frenk} et~al.}{1988}]{frenk1988}
{Frenk} C.~S.,  {White} S.~D.~M.,  {Davis} M.,   {Efstathiou} G.,  1988,
  \mn@doi [\apj] {10.1086/166213}, \href
  {http://adsabs.harvard.edu/abs/1988ApJ...327..507F} {327, 507}

\bibitem[\protect\citeauthoryear{Fry et~al.,}{Fry et~al.}{2015}]{Fry:2015rta}
Fry A.~B.,  et~al., 2015, \mn@doi [Mon. Not. Roy. Astron. Soc.]
  {10.1093/mnras/stv1330}, 452, 1468

\bibitem[\protect\citeauthoryear{{Garrison-Kimmel}, {Rocha}, {Boylan-Kolchin},
  {Bullock}  \& {Lally}}{{Garrison-Kimmel} et~al.}{2013}]{garrison-kimmel2013}
{Garrison-Kimmel} S.,  {Rocha} M.,  {Boylan-Kolchin} M.,  {Bullock} J.~S.,
  {Lally} J.,  2013, \mn@doi [\mnras] {10.1093/mnras/stt984}, \href
  {http://adsabs.harvard.edu/abs/2013MNRAS.433.3539G} {433, 3539}

\bibitem[\protect\citeauthoryear{{Hayashi}, {Navarro}  \& {Springel}}{{Hayashi}
  et~al.}{2007}]{hayashi2007}
{Hayashi} E.,  {Navarro} J.~F.,   {Springel} V.,  2007, \mn@doi [\mnras]
  {10.1111/j.1365-2966.2007.11599.x}, \href
  {http://adsabs.harvard.edu/abs/2007MNRAS.377...50H} {377, 50}

\bibitem[\protect\citeauthoryear{{Hernquist}}{{Hernquist}}{1990}]{hernquist-1990}
{Hernquist} L.,  1990, \mn@doi [\apj] {10.1086/168845}, \href
  {http://adsabs.harvard.edu/abs/1990ApJ...356..359H} {356, 359}

\bibitem[\protect\citeauthoryear{{Jing} \& {Suto}}{{Jing} \&
  {Suto}}{2002}]{suto2002}
{Jing} Y.~P.,  {Suto} Y.,  2002, \mn@doi [\apj] {10.1086/341065}, \href
  {http://adsabs.harvard.edu/abs/2002ApJ...574..538J} {574, 538}

\bibitem[\protect\citeauthoryear{{Kamada}, {Kaplinghat}, {Pace}  \&
  {Yu}}{{Kamada} et~al.}{2017}]{2017PhRvL.119k1102K}
{Kamada} A.,  {Kaplinghat} M.,  {Pace} A.~B.,   {Yu} H.-B.,  2017, \mn@doi
  [Physical Review Letters] {10.1103/PhysRevLett.119.111102}, \href
  {http://adsabs.harvard.edu/abs/2017PhRvL.119k1102K} {119, 111102}

\bibitem[\protect\citeauthoryear{{Kaplinghat}, {Keeley}, {Linden}  \&
  {Yu}}{{Kaplinghat} et~al.}{2014}]{kaplinghat2014a}
{Kaplinghat} M.,  {Keeley} R.~E.,  {Linden} T.,   {Yu} H.-B.,  2014, \mn@doi
  [Physical Review Letters] {10.1103/PhysRevLett.113.021302}, \href
  {http://adsabs.harvard.edu/abs/2014PhRvL.113b1302K} {113, 021302}

\bibitem[\protect\citeauthoryear{{Kaplinghat}, {Tulin}  \& {Yu}}{{Kaplinghat}
  et~al.}{2016}]{kaplinghat2015}
{Kaplinghat} M.,  {Tulin} S.,   {Yu} H.-B.,  2016, \mn@doi [Physical Review
  Letters] {10.1103/PhysRevLett.116.041302}, \href
  {http://adsabs.harvard.edu/abs/2016PhRvL.116d1302K} {116, 041302}

\bibitem[\protect\citeauthoryear{{Kazantzidis}, {Abadi}  \&
  {Navarro}}{{Kazantzidis} et~al.}{2010}]{kazantzidis2010}
{Kazantzidis} S.,  {Abadi} M.~G.,   {Navarro} J.~F.,  2010, \mn@doi [\apjl]
  {10.1088/2041-8205/720/1/L62}, \href
  {http://adsabs.harvard.edu/abs/2010ApJ...720L..62K} {720, L62}

\bibitem[\protect\citeauthoryear{{Koda} \& {Shapiro}}{{Koda} \&
  {Shapiro}}{2011}]{koda2011}
{Koda} J.,  {Shapiro} P.~R.,  2011, \mn@doi [\mnras]
  {10.1111/j.1365-2966.2011.18684.x}, \href
  {http://adsabs.harvard.edu/abs/2011MNRAS.415.1125K} {415, 1125}

\bibitem[\protect\citeauthoryear{{Koposov}, {Rix}  \& {Hogg}}{{Koposov}
  et~al.}{2010}]{koposov2010}
{Koposov} S.~E.,  {Rix} H.-W.,   {Hogg} D.~W.,  2010, \mn@doi [\apj]
  {10.1088/0004-637X/712/1/260}, \href
  {http://adsabs.harvard.edu/abs/2010ApJ...712..260K} {712, 260}

\bibitem[\protect\citeauthoryear{{Kuhlen}, {Diemand}  \& {Madau}}{{Kuhlen}
  et~al.}{2007}]{kuhlen2007}
{Kuhlen} M.,  {Diemand} J.,   {Madau} P.,  2007, \mn@doi [\apj]
  {10.1086/522878}, \href {http://adsabs.harvard.edu/abs/2007ApJ...671.1135K}
  {671, 1135}

\bibitem[\protect\citeauthoryear{{K{\"u}pper}, {Balbinot}, {Bonaca},
  {Johnston}, {Hogg}, {Kroupa}  \& {Santiago}}{{K{\"u}pper}
  et~al.}{2015}]{kupper2015}
{K{\"u}pper} A.~H.~W.,  {Balbinot} E.,  {Bonaca} A.,  {Johnston} K.~V.,  {Hogg}
  D.~W.,  {Kroupa} P.,   {Santiago} B.~X.,  2015, \mn@doi [\apj]
  {10.1088/0004-637X/803/2/80}, \href
  {http://adsabs.harvard.edu/abs/2015ApJ...803...80K} {803, 80}

\bibitem[\protect\citeauthoryear{Kuzio~de Naray, Martinez, Bullock  \&
  Kaplinghat}{Kuzio~de Naray et~al.}{2010}]{deNaray:2009xj}
Kuzio~de Naray R.,  Martinez G.~D.,  Bullock J.~S.,   Kaplinghat M.,  2010,
  \mn@doi [Astrophys. J.] {10.1088/2041-8205/710/2/L161}, 710, L161

\bibitem[\protect\citeauthoryear{{Lelli}, {McGaugh}  \& {Schombert}}{{Lelli}
  et~al.}{2016}]{Lelli}
{Lelli} F.,  {McGaugh} S.~S.,   {Schombert} J.~M.,  2016, \mn@doi [\apjl]
  {10.3847/2041-8205/816/1/L14}, \href
  {http://adsabs.harvard.edu/abs/2016ApJ...816L..14L} {816, L14}

\bibitem[\protect\citeauthoryear{{McMillan}}{{McMillan}}{2011}]{mcmillan}
{McMillan} P.~J.,  2011, \mn@doi [\mnras] {10.1111/j.1365-2966.2011.18564.x},
  \href {http://adsabs.harvard.edu/abs/2011MNRAS.414.2446M} {414, 2446}

\bibitem[\protect\citeauthoryear{{Miyamoto} \& {Nagai}}{{Miyamoto} \&
  {Nagai}}{1975}]{Miyamoto-Nagai}
{Miyamoto} M.,  {Nagai} R.,  1975, \pasj, \href
  {http://adsabs.harvard.edu/abs/1975PASJ...27..533M} {27, 533}

\bibitem[\protect\citeauthoryear{{Navarro}, {Frenk}  \& {White}}{{Navarro}
  et~al.}{1997}]{NFW}
{Navarro} J.~F.,  {Frenk} C.~S.,   {White} S.~D.~M.,  1997, \apj, \href
  {http://adsabs.harvard.edu/abs/1997ApJ...490..493N} {490, 493}

\bibitem[\protect\citeauthoryear{{Navarro} et~al.,}{{Navarro}
  et~al.}{2010}]{Navarro2008}
{Navarro} J.~F.,  et~al., 2010, \mn@doi [\mnras]
  {10.1111/j.1365-2966.2009.15878.x}, \href
  {http://adsabs.harvard.edu/abs/2010MNRAS.402...21N} {402, 21}

\bibitem[\protect\citeauthoryear{{Oman} et~al.,}{{Oman} et~al.}{2015}]{Oman}
{Oman} K.~A.,  et~al., 2015, \mn@doi [\mnras] {10.1093/mnras/stv1504}, \href
  {http://adsabs.harvard.edu/abs/2015MNRAS.452.3650O} {452, 3650}

\bibitem[\protect\citeauthoryear{{Pearson}, {K{\"u}pper}, {Johnston}  \&
  {Price-Whelan}}{{Pearson} et~al.}{2015}]{pearson2015}
{Pearson} S.,  {K{\"u}pper} A.~H.~W.,  {Johnston} K.~V.,   {Price-Whelan}
  A.~M.,  2015, \mn@doi [\apj] {10.1088/0004-637X/799/1/28}, \href
  {http://adsabs.harvard.edu/abs/2015ApJ...799...28P} {799, 28}

\bibitem[\protect\citeauthoryear{{Peter}, {Rocha}, {Bullock}  \&
  {Kaplinghat}}{{Peter} et~al.}{2013}]{A.Peter}
{Peter} A.~H.~G.,  {Rocha} M.,  {Bullock} J.~S.,   {Kaplinghat} M.,  2013,
  \mn@doi [\mnras] {10.1093/mnras/sts535}, \href
  {http://adsabs.harvard.edu/abs/2013MNRAS.430..105P} {430, 105}

\bibitem[\protect\citeauthoryear{{Planck Collaboration} et~al.,}{{Planck
  Collaboration} et~al.}{2014}]{planck2014}
{Planck Collaboration} et~al., 2014, \mn@doi [\aap]
  {10.1051/0004-6361/201321591}, \href
  {http://adsabs.harvard.edu/abs/2014A%26A...571A..16P} {571, A16}

\bibitem[\protect\citeauthoryear{{Power}, {Navarro}, {Jenkins}, {Frenk},
  {White}, {Springel}, {Stadel}  \& {Quinn}}{{Power} et~al.}{2003}]{power2003}
{Power} C.,  {Navarro} J.~F.,  {Jenkins} A.,  {Frenk} C.~S.,  {White} S.~D.~M.,
   {Springel} V.,  {Stadel} J.,   {Quinn} T.,  2003, \mn@doi [\mnras]
  {10.1046/j.1365-8711.2003.05925.x}, \href
  {http://adsabs.harvard.edu/abs/2003MNRAS.338...14P} {338, 14}

\bibitem[\protect\citeauthoryear{{Robertson} et~al.,}{{Robertson}
  et~al.}{2017}]{2017arXiv171109096R}
{Robertson} A.,  et~al., 2017, preprint, \href
  {http://adsabs.harvard.edu/abs/2017arXiv171109096R} {} (\mn@eprint {arXiv}
  {1711.09096})

\bibitem[\protect\citeauthoryear{{Robles} et~al.,}{{Robles}
  et~al.}{2017}]{2017MNRAS.472.2945R}
{Robles} V.~H.,  et~al., 2017, \mn@doi [\mnras] {10.1093/mnras/stx2253}, \href
  {http://adsabs.harvard.edu/abs/2017MNRAS.472.2945R} {472, 2945}

\bibitem[\protect\citeauthoryear{{Rocha}, {Peter}, {Bullock}, {Kaplinghat},
  {Garrison-Kimmel}, {O{\~n}orbe}  \& {Moustakas}}{{Rocha}
  et~al.}{2013}]{rocha2013}
{Rocha} M.,  {Peter} A.~H.~G.,  {Bullock} J.~S.,  {Kaplinghat} M.,
  {Garrison-Kimmel} S.,  {O{\~n}orbe} J.,   {Moustakas} L.~A.,  2013, \mn@doi
  [\mnras] {10.1093/mnras/sts514}, \href
  {http://adsabs.harvard.edu/abs/2013MNRAS.430...81R} {430, 81}

\bibitem[\protect\citeauthoryear{{Spergel} \& {Steinhardt}}{{Spergel} \&
  {Steinhardt}}{2000}]{spergel1999}
{Spergel} D.~N.,  {Steinhardt} P.~J.,  2000, \mn@doi [Physical Review Letters]
  {10.1103/PhysRevLett.84.3760}, \href
  {http://adsabs.harvard.edu/abs/2000PhRvL..84.3760S} {84, 3760}

\bibitem[\protect\citeauthoryear{{Springel}}{{Springel}}{2005}]{springel2005}
{Springel} V.,  2005, \mn@doi [\mnras] {10.1111/j.1365-2966.2005.09655.x},
  \href {http://adsabs.harvard.edu/abs/2005MNRAS.364.1105S} {364, 1105}

\bibitem[\protect\citeauthoryear{{Springel}}{{Springel}}{2010}]{Springel2010}
{Springel} V.,  2010, \mn@doi [\mnras] {10.1111/j.1365-2966.2009.15715.x},
  \href {http://adsabs.harvard.edu/abs/2010MNRAS.401..791S} {401, 791}

\bibitem[\protect\citeauthoryear{Springel, Frenk  \& White}{Springel
  et~al.}{2006}]{Springel:2006vs}
Springel V.,  Frenk C.~S.,   White S. D.~M.,  2006, \mn@doi [Nature]
  {10.1038/nature04805}, 440, 1137

\bibitem[\protect\citeauthoryear{Springel et~al.,}{Springel
  et~al.}{2008}]{springel2008}
Springel V.,  et~al., 2008, \mn@doi [Mon. Not. Roy. Astron. Soc.]
  {10.1111/j.1365-2966.2008.14066.x}, 391, 1685

\bibitem[\protect\citeauthoryear{{Tissera}, {White}, {Pedrosa}  \&
  {Scannapieco}}{{Tissera} et~al.}{2010}]{tissera2010}
{Tissera} P.~B.,  {White} S.~D.~M.,  {Pedrosa} S.,   {Scannapieco} C.,  2010,
  \mn@doi [\mnras] {10.1111/j.1365-2966.2010.16777.x}, \href
  {http://adsabs.harvard.edu/abs/2010MNRAS.406..922T} {406, 922}

\bibitem[\protect\citeauthoryear{{Tollet} et~al.,}{{Tollet}
  et~al.}{2016}]{tollet2016}
{Tollet} E.,  et~al., 2016, \mn@doi [\mnras] {10.1093/mnras/stv2856}, \href
  {http://adsabs.harvard.edu/abs/2016MNRAS.456.3542T} {456, 3542}

\bibitem[\protect\citeauthoryear{Trujillo-Gomez, Klypin, Primack  \&
  Romanowsky}{Trujillo-Gomez et~al.}{2011}]{TrujilloGomez:2010yh}
Trujillo-Gomez S.,  Klypin A.,  Primack J.,   Romanowsky A.~J.,  2011, \mn@doi
  [Astrophys. J.] {10.1088/0004-637X/742/1/16}, 742, 16

\bibitem[\protect\citeauthoryear{{Tulin} \& {Yu}}{{Tulin} \&
  {Yu}}{2017}]{2017arXiv170502358T}
{Tulin} S.,  {Yu} H.-B.,  2017, preprint, \href
  {http://adsabs.harvard.edu/abs/2017arXiv170502358T} {} (\mn@eprint {arXiv}
  {1705.02358})

\bibitem[\protect\citeauthoryear{{Vera-Ciro}, {Sales}, {Helmi}, {Frenk},
  {Navarro}, {Springel}, {Vogelsberger}  \& {White}}{{Vera-Ciro}
  et~al.}{2011}]{Vera-Ciro1}
{Vera-Ciro} C.~A.,  {Sales} L.~V.,  {Helmi} A.,  {Frenk} C.~S.,  {Navarro}
  J.~F.,  {Springel} V.,  {Vogelsberger} M.,   {White} S.~D.~M.,  2011, \mn@doi
  [\mnras] {10.1111/j.1365-2966.2011.19134.x}, \href
  {http://adsabs.harvard.edu/abs/2011MNRAS.416.1377V} {416, 1377}

\bibitem[\protect\citeauthoryear{{Vogelsberger}, {Zavala}  \&
  {Loeb}}{{Vogelsberger} et~al.}{2012}]{Volgesberger2012}
{Vogelsberger} M.,  {Zavala} J.,   {Loeb} A.,  2012, \mn@doi [\mnras]
  {10.1111/j.1365-2966.2012.21182.x}, \href
  {http://adsabs.harvard.edu/abs/2012MNRAS.423.3740V} {423, 3740}

\bibitem[\protect\citeauthoryear{{Vogelsberger} et~al.,}{{Vogelsberger}
  et~al.}{2014a}]{2014MNRAS.444.1518V}
{Vogelsberger} M.,  et~al., 2014a, \mn@doi [\mnras] {10.1093/mnras/stu1536},
  \href {http://adsabs.harvard.edu/abs/2014MNRAS.444.1518V} {444, 1518}

\bibitem[\protect\citeauthoryear{{Vogelsberger}, {Zavala}, {Simpson}  \&
  {Jenkins}}{{Vogelsberger} et~al.}{2014b}]{Vogelsberger_2014}
{Vogelsberger} M.,  {Zavala} J.,  {Simpson} C.,   {Jenkins} A.,  2014b, \mn@doi
  [\mnras] {10.1093/mnras/stu1713}, \href
  {http://adsabs.harvard.edu/abs/2014MNRAS.444.3684V} {444, 3684}

\bibitem[\protect\citeauthoryear{Vogelsberger et~al.,}{Vogelsberger
  et~al.}{2014c}]{Vogelsberger:2014kha}
Vogelsberger M.,  et~al., 2014c, \mn@doi [Nature] {10.1038/nature13316}, 509,
  177

\bibitem[\protect\citeauthoryear{{Vogelsberger}, {Zavala}, {Cyr-Racine},
  {Pfrommer}, {Bringmann}  \& {Sigurdson}}{{Vogelsberger}
  et~al.}{2016}]{2016MNRAS.460.1399V}
{Vogelsberger} M.,  {Zavala} J.,  {Cyr-Racine} F.-Y.,  {Pfrommer} C.,
  {Bringmann} T.,   {Sigurdson} K.,  2016, \mn@doi [\mnras]
  {10.1093/mnras/stw1076}, \href
  {http://adsabs.harvard.edu/abs/2016MNRAS.460.1399V} {460, 1399}

\bibitem[\protect\citeauthoryear{{Yoshida}, {Springel}, {White}  \&
  {Tormen}}{{Yoshida} et~al.}{2000}]{yoshida2000}
{Yoshida} N.,  {Springel} V.,  {White} S.~D.~M.,   {Tormen} G.,  2000, \mn@doi
  [\apjl] {10.1086/317306}, \href
  {http://adsabs.harvard.edu/abs/2000ApJ...544L..87Y} {544, L87}

\bibitem[\protect\citeauthoryear{{Zavala}, {Vogelsberger}  \&
  {Walker}}{{Zavala} et~al.}{2013}]{zavala2013}
{Zavala} J.,  {Vogelsberger} M.,   {Walker} M.~G.,  2013, \mn@doi [\mnras]
  {10.1093/mnrasl/sls053}, \href
  {http://adsabs.harvard.edu/abs/2013MNRAS.431L..20Z} {431, L20}

\makeatother
\end{thebibliography}
\appendix
\section{Convergence test for the halo shape algorithm}\label{sec:A1}
\begin{figure}
\includegraphics[width=\columnwidth]{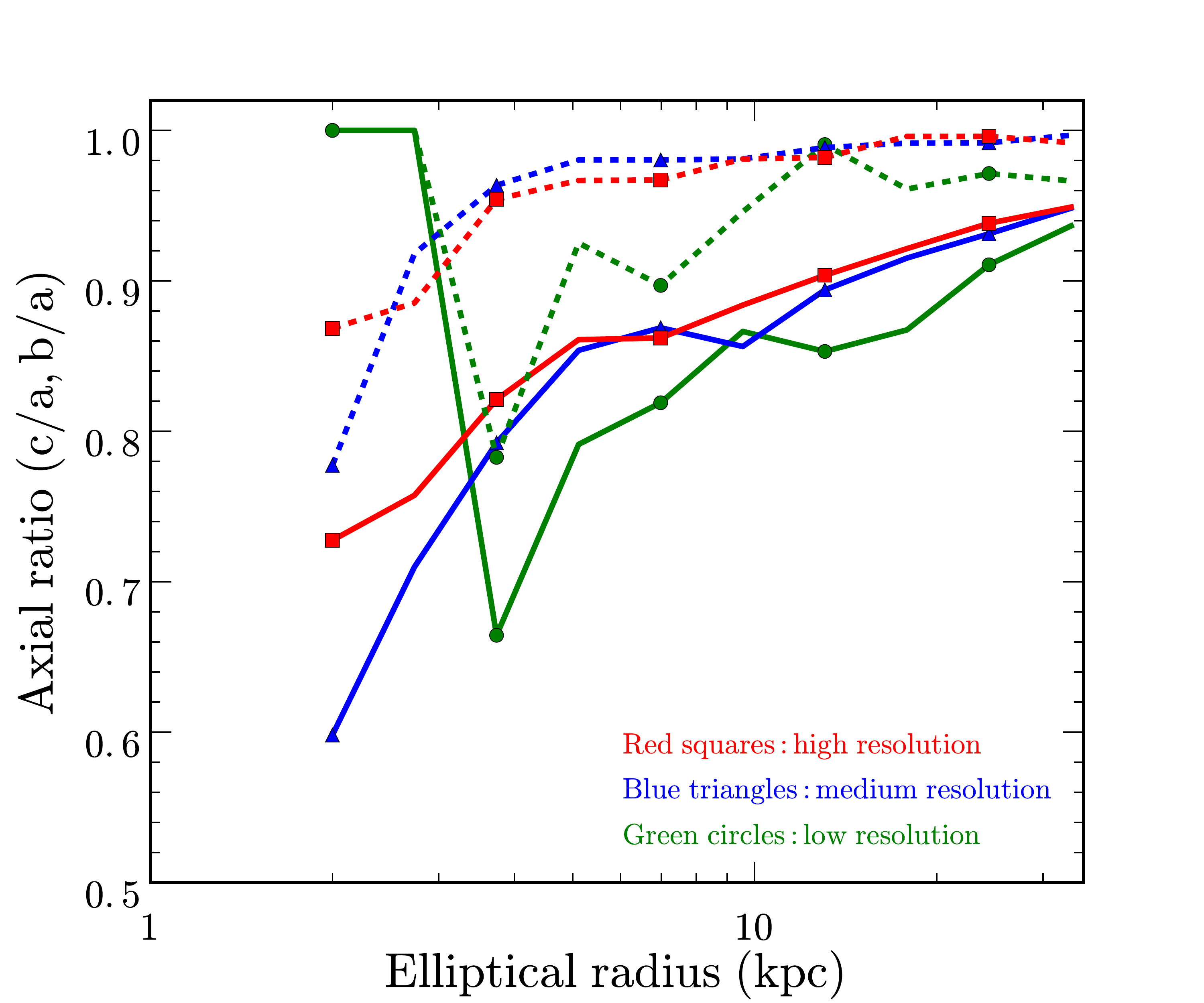}
\caption{Ratio of minor- (solid) and intermediate-to-major (dashed) axes for the convergence test runs.
}
\label{fig:conv_test}
\end{figure}

To evaluate accuracy of the shape measurement, we follow~\citet{Vera-Ciro1} to determine the convergence radius where shape measurements are robust in our simulations. 
We consider the convergence radius $r_{\rm conv}$~\citep{power2003,Navarro2008},
\begin{equation}
\nonumber\kappa (r_{\rm conv}) \equiv \frac{t_{\rm relax}}{t_{\rm cir}(r_{200})} = \frac{\sqrt{200}}{8}
\frac{N(r_{\rm conv})}{\ln{N(r_{\rm conv})}}
\left(\frac{\overline{\rho}(r_{\rm conv})}{\rho_{\rm crit}}\right)^{-1/2} 
\end{equation}
where $t_{\rm relax}$ is the two-body relaxation time scale due to gravitational encounters, $t_{\rm cir}$ is the circular orbit timescale at $r_{200}$, $N$ is number of DM particles and $\overline{\rho}$ is the average density inside the convergence radius. We take $\kappa (r_{\rm conv})=7$ as in~\citet{Vera-Ciro1}. In addition, we require $\sim 70\%$ of the particles to be inside the virial sphere and at least $2000$ DM particles inside of convergence radius. 

The first requirement is satisfied if we choose a large cutoff radius in the {\sc SpherIC} code, at which the density profile transits from an NFW one to an exponentially decaying one to avoid the divergence of the mass. Then we run three simulations with different values of the DM particle number and the gravitational softening length as a convergence test with details summarized in Table~\ref{table:IC_param}. The convergence radius decreases with increasing the number of total DM particles. Thus, to probe the shape of the inner halo, down to few ${\rm kpc}$, we need at least $2$ million particles in simulations. Fig.~\ref{fig:conv_test} shows the $b/a$ (top) and $c/a$ (bottom) profiles for different resolutions, we take a static MN potential as stellar disc similar to Sec.~\ref{sec:halo-disc}, with a MW-sized halo with $M_{200} \simeq\ 2.6\times 10^{12}\ M_{\odot}$. We see that the convergence improves when $N_{\rm tot}$ increases. We take high-resolution run in the results presented in Sec.~\ref{sec:halo-disc} and~\ref{sec:MW}.

\begin{table}
\begin{center}
\resizebox{\columnwidth}{!}{
\begin{tabular}{cccccc}
\hline
$\text{Resolution Level}$ & $\text{m}_{\rm p}\ (M_{\odot})$ & $\text{N}_{\rm tot}$ & $\epsilon\ (\text{pc})$ & $\text{r}_{\rm conv}\ (\text{kpc})$ & $\text{N}_{\rm conv}$\\
\hline
\hline
Low & $2.18\times 10^8$ &$1.6 \times 10^{5}$ & $500$ & 6.08 & 1190\\
Intermediate & $3.97\times 10^6$ &$8.8\times 10^5$ & $250$ & 2.99  & 1813  \\ 
High & $1.31\times 10^6$ &$2\times 10^6$ & $125$  & 2.18  & 2192\\
\hline
\end{tabular}
}
\end{center}
\caption{Summary  of simulations for convergence test. $\text{m}_{\rm p}$ is mass of each
particle in each simulation, $N_{\rm tot}$ is total 
number of particles, $\epsilon$ is gravitational softening length,
$r_{\rm conv}$ is convergence radius, and $N_{\rm conv}$ is number of particles inside 
$r_{\rm conv}$. 
}
\label{table:IC_param}
\end{table}

\end{document}